\documentclass[aps,pre,twocolumn,superscriptaddress]{revtex4}
\usepackage{amsmath, epsfig}

\renewcommand{\vr}{{\bf{r}}}
\newcommand{\vR}{{\bf{R}}}

\newcommand{\vk}{{\bf{k}}}
\newcommand{\vQ}{{\bf Q}}
\newcommand{\vq}{{\bf q}}
\newcommand{\unit}{{\bf 1}}
\newcommand{\tht}{\theta}
\newcommand{\kpp}{\kappa}
\newcommand{\la}{\lambdabar}
\newcommand{\sdif}{{\sigma_\mathrm{diff}}}

\begin{document}

\title{Wave packet autocorrelation functions for quantum hard-disk and
  hard-sphere billiards in the high-energy, diffraction regime}

\author{Arseni Goussev} \affiliation{Institut f\"ur Theoretische Physik,
  Universit\"at Regensburg, 93040 Regensburg, Germany}

\author{J. R. Dorfman} \affiliation{Institute for Physical Science and
  Technology and Department of Physics, University of Maryland,\\
  College Park, MD 20742 USA}

\date{\today}

\begin{abstract}
We consider the time evolution of a wave packet representing a quantum
particle moving in a geometrically open billiard that consists of a
number of fixed hard-disk or hard-sphere scatterers. Using the
technique of multiple collision expansions we provide a
first-principle analytical calculation of the time-dependent
autocorrelation function for the wave packet in the high-energy
diffraction regime, in which the particle's de Broglie wave length,
while being small compared to the size of the scatterers, is large
enough to prevent the formation of geometric shadow over distances of
the order of the particle's free flight path. The hard-disk or
hard-sphere scattering system must be sufficiently dilute in order for
this high-energy diffraction regime to be achievable. Apart from the
overall exponential decay, the autocorrelation function exhibits a
generally complicated sequence of relatively strong peaks corresponding
to partial revivals of the wave packet. Both the exponential decay (or
escape) rate and the revival peak structure are
predominantly determined by the underlying classical dynamics. A
relation between the escape rate, and the Lyapunov exponents and
Kolmogorov-Sinai entropy of the counterpart classical system,
previously known for hard-disk billiards, is strengthened by
generalization to three spatial dimensions. The results of the quantum
mechanical calculation of the time-dependent autocorrelation function
agree with predictions of the semiclassical periodic
orbit theory.
\end{abstract}

\maketitle

\section{Introduction}

One of the main objectives of the field of quantum chaos is to study
the quantum mechanics of systems that are chaotic in the classical
limit, and to find what properties of the classical systems appear
crucial for the quantum description. There are some common approaches
to the problem. One can explore the energy spectra of quantum analogs
to classically chaotic, bounded systems
\cite{haake,stock,berry,sieb,smil-1}, or investigate scattering
resonances in the complex energy plane for unbounded quantum systems
with chaotic classical repellers
\cite{gasp-II,gasp-III,alons,wirz-1,wirz-2,cvit}.  Interesting results
relating the energy spectra of closed systems to the scattering
resonances of the complementary open systems (defined as the region
``outside'' the closed system) are also known \cite{smil-2}.  Another
approach focuses on the time-dependent description, which involves
such quantities as the quantum state autocorrelation functions
\cite{peres-1,hell-1,hell-2,toms,robin,us} and the Loschmidt echo, or
fidelity \cite{peres-2,pros,cucc}. In this paper we continue the
second approach and address the time evolution of the autocorrelation
function for quantum particles traveling in hard-disk and hard-sphere
billiards in two and three spatial dimensions respectively.

We have previously studied the short time dynamics of a small
Gaussian wave packet in arrays of hard-disk scatterers \cite{us}. The
wave packet was considered to be much smaller than the disk radius,
and its evolution was limited to times shorter than the Ehrenfest time
\cite{haake}, at which time the wave packet size becomes comparable with
the scatterer size. The goal of the present paper is to extend the
previous results by investigating the wave packet dynamics in the
limit of long times, much longer than the Ehrenfest time. Here, we
show that substantial progress can be made for hard-ball, (hard disk
or sphere) scattering systems containing only a small number of
scatterers, and for wave packets in the diffraction regime, to be
further specified below.

In a series of seminal papers \cite{gasp-I,gasp-II,gasp-III} Gaspard
and Rice studied classical, semiclassical and quantum properties of a
three-disk scattering system, in which a point particle moves in two
dimensions in the presence of three fixed, impenetrable disk
scatterers. The disks have the radius $a$, and their centers are
placed at the vertices of an equilateral triangle of side $R$.
Below we will refer to this scattering system as to the three-disk
equilateral billiard to distinguish it from other three-disk
billiards, in which the scatterer centers form isosceles or generic
triangles. The latter scattering systems will be referred as to the
isosceles billiard and the generic triangle billiard, respectively.  The
classical version of the three-disk system is known to have a fractal
repeller with the positive Kolmogorov-Sinai entropy \cite{gasp-I}. The
quantum dynamics of a particle in the three-disk billiard is largely
determined by scattering resonances, whose distribution in the complex
energy plane reflects the chaotic properties of the corresponding
classical system \cite{gasp-II,gasp-III}.

In this paper we address the time evolution of initially localized
wave packets in such hard-disk scattering systems as the two-disk
billiard and the three-disk equilateral, isosceles and generic
triangular billiards. We also extend our calculations to include other
hard-sphere scattering systems in three spatial dimensions. As a tool
for quantifying the wave packet dynamics we consider the
autocorrelation function $C(t)$ defined as the overlap
\begin{equation}
C(t) = \left| \langle \phi_0 | \phi_t \rangle \right|^2,
\label{intr-1}
\end{equation}
where $|\phi_0 \rangle$ is the initial quantum state of a billiard,
and $|\phi_t \rangle$ is the quantum state obtained from the initial
one by the time evolution through time $t$ under the Hamiltonian $H$:
\begin{equation}
| \phi_t \rangle = e^{-i Ht/\hbar} | \phi_0 \rangle.
\label{intr-2}
\end{equation}
We construct the quantum propagator, $e^{-i Ht/\hbar}$, for a dilute
hard-disk billiard using the multiple collision expansion technique
\cite{wats,joach}. Then, we apply this propagator to an initially
localized wave packet $|\phi_0 \rangle$, corresponding to a highly-energetic
particle traveling through the system, and calculate the
autocorrelation function $C(t)$. The following two assumptions allow
analytical treatment of the problem: (i) the scattering system is
considered to be dilute, and (ii) the quantum particle is restricted
to have a short de Broglie wave length. (We will give further details
below.) These assumptions limit the validity of our results for the
autocorrelation function to a certain time range. The latter can be
quite long: it is bounded from below by the Ehrenfest time of the
system, and typically extends over a large number of Ehrenfest times.
Therefore, this work should be considered as an attempt to
analytically describe the dynamics of initially localized wave packets
over long times during which the quantum particle explores all the
scattering system, and its wave function becomes spatially extended.

It was pointed out by Gaspard {et al.} \cite{alons} that the time
evolution of a wave packet in a hard-disk scattering system exhibits
an overall exponential damping controlled by the location of the
scattering resonances in the complex energy plane. As a result of this
damping one expects the wave packet autocorrelation function, $C(t)$,
to decay in an exponential-like manner as well. We explicitly
calculate the autocorrelation function for wave packets in various
hard-disk billiards and investigate the structure of its time decay.
We find that the decay consists of a sequence of sharp peaks with the
exponentially decreasing envelope. The peaks reflect the phenomenon of
the wave packet partial reconstruction due to interference of
different classical periodic orbits in the billiard. The maxima of the
peaks occur at times at which the classical particle, having the
momentum equal to the average momentum of the wave packet, would
return to its initial phase-space point. Although the long-time
autocorrelation function peaks have the same origin as the ones in the
short-time autocorrelation function \cite{hell-2,us}, the peaks at
longer times result from the constructive interference of a large
number of classical trajectories whereas only a single classical path
is responsible for the wave packet reconstruction during times shorter
than the Ehrenfest time.

The envelope of the autocorrelation function for a two-dimensional
hard-disk billiard decays exponentially with time, $e^{-\gamma t}$, with
the rate $\gamma$ approximately equal to the difference of the mean
positive Lyapunov exponent, $\lambda$, and twice the Kolmogorov-Sinai (KS)
entropy per unit time, $h_\mathrm{KS}$, of the repeller of the
corresponding classical system, i.e.  $\gamma \approx \lambda-2h_\mathrm{KS}$. This
decay rate is nothing but the quantum escape rate originally obtained
by Gaspard and Rice \cite{gasp-II}. Our generalization to three
spatial dimensions shows that the envelope of the wave packet
autocorrelation function also decays exponentially, $e^{-\gamma_\mathrm{3D}
  t}$, with the decay rate $\gamma_\mathrm{3D} = \lambda_1 + \lambda_2 -
2h_\mathrm{KS}$, where the place of $\lambda$ in the expression for $\gamma$ is
taken by the sum of the two mean positive Lyapunov exponents $\lambda_1$ and
$\lambda_2$ of the classical hard-sphere system.

The paper is organized as follows. In Section~II we calculate the
autocorrelation function, $C(t)$, for wave packets in two-disk and
various three-disk scattering systems using the method of the multiple
collision expansion. We show that the structure of $C(t)$ is generally
quite intricate, and we provide explanations of its main features.
Section~III presents an alternative description of the autocorrelation
function decay based on the semiclassical arguments. While the
semi-classical method described here does not provide as much detail
about the autocorrelation functions as do the multiple scattering
methods, it does provide information about the relative strengths of
the peaks and their location. Moreover this method allows us to obtain
information about more complicated systems that are difficult to treat
using more exact methods. For example, in this section we use the
semi-classical method to treat generic three disk scatterers, and to
generalize our methods to three spatial dimensions to describe the
wave packet dynamics in two-, three- and four-sphere scattering
systems. Our conclusions and discussions are contained in the
Chapter~IV of the paper.

\section{Hard-disk scattering systems}

\subsection{General formulation}

We consider a particle of mass $m$ placed among a collection of $N$
fixed hard-disk scatterers of radius $a$ centered at position vectors
$\vR_j$, with $j=1,2,\ldots,N$.  The Hamiltonian of the system can be
written as
\begin{equation}
H = H_0  + \sum_{j=1}^N V_j,
\label{2.1}
\end{equation}
where $H_0$ is the free particle Hamiltonian, and the hard-disk
scatterer potentials are given by
\begin{equation}
V_j(\vr)= \left\{
\begin{array}{ll}
+\infty \;\;\; &\mathrm{for} \;\; |\vr-\vR_j|<a,\\ 0 &\mathrm{for}
\;\; |\vr-\vR_j| \geq a.
\end{array} \right.
\label{2.2}
\end{equation}
The time-domain propagator, $G(t)=e^{-iHt/\hbar}$, satisfying the
Schr\"odinger equation with the Hamiltonian $H$, evolves an initial
quantum state $|\phi_0 \rangle$ in time in accordance with
Eq.~(\ref{intr-2}). The corresponding energy-domain propagator is
obtained by means of the positive time Fourier transform
\begin{equation}
G(E) = \frac{1}{i\hbar} \int_0^{+\infty} dt
e^{i(E+i0)t/\hbar} G(t) = \frac{1}{E-H+i0}.
\label{2.3}
\end{equation}
The inverse transform is given by
\begin{equation}
G(t) = \frac{i}{2\pi} \int_{-\infty}^{+\infty} dE e^{-iEt/\hbar} G(E),
\label{2.4}
\end{equation}
where we consider time $t$ to be strictly positive. Although it is a
formidable problem to construct the time-domain propagator directly,
one can employ the method of multiple collisions \cite{wats,joach} to
obtain a series expansion for the energy-dependent propagator.

The multiple collision expansion of the energy-dependent propagator
represents $G(E)$ as an infinite sum over possible collision sequences
that the quantum particle can undergo:
\begin{equation}
\begin{split}
  G = &G_0 + \sum_j G_0 T_j G_0 + \sum_j \sum_{k \neq j} G_0 T_j G_0 T_k G_0\\ &+
  \sum_j \sum_{k \neq j} \sum_{l \neq k} G_0 T_j G_0 T_k G_0 T_l G_0 + \ldots,
\end{split}
\label{2.5}
\end{equation}
where $G_0 = (E-H_0+i0)^{-1}$ is the free particle propagator in the
energy domain, and the binary collision operator, $T_j$, also known as
the T-matrix, is defined by the multiple collision expansion,
Eq.~(\ref{2.5}), written for a system containing only the $j^{\,
  \mathrm{th}}$ scatterer, {\it i.e.}
\begin{equation}
G_j \equiv \frac{1}{E-H_0-V_j+i0} = G_0 + G_0 T_j G_0.
\label{2.6}
\end{equation}
Hereafter all $G$ and $T$ operators are given in the energy domain
unless the time dependence is specified explicitly. Here, $G_j$ is the
propagator for a particle moving in two dimensional space with only
one scatterer located at point $\vR_j$.

The momentum-space matrix elements of the binary collision operator
for the hard-disk potential given by Eq.~(\ref{2.2}) were calculated
by Correia \cite{corr}. For a particle of energy $E=\hbar^2
\kpp^2/2m$, where $\kpp$ is the magnitude of the particle's
wave vector, the matrix element $\langle \vk | T_j | \vk' \rangle$ of
the binary collision operator, relating two generally different
momentum states $\vk$ and $\vk'$, is given by
\begin{equation}
\begin{split}
  &\langle \vk | T_j | \vk' \rangle = 2\pi a \frac{\hbar^2}{2m}
  e^{-i(\vk-\vk')\vR_j}\sum_{l=-\infty}^{+\infty} e^{i l(\tht_k-\tht_{k'})}\\ &\times
  \left\{ \frac{(k')^2-\kpp^2}{k^2-(k')^2} \left[ k' J_l(ka)
      J_{l-1}(k'a) - k J_{l-1}(ka) J_l(k'a) \right] \right.\\
  &\phantom{++} \left. + k' J_l(ka) J_{l-1}(k'a) - \kpp J_l(ka)
    J_l(k'a) \frac{H_{l-1}^{(1)}(\kpp a)}{H_l^{(1)}(\kpp a)} \right\},
\end{split}
\label{2.7}
\end{equation}
where $(k,\tht_k)$ and $(k',\tht_{k'})$ are the polar coordinates of
the wave vectors $\vk$ and $\vk'$ respectively, $J_l$ is the Bessel
function of the first kind of order $l$, $H_l^{(1)}$ is the Hankel
function of the first kind of order $l$. Hereafter, the following
normalization conditions are adopted: $\langle \vr | \vr' \rangle =
\delta(\vr-\vr')$, $\langle \vk | \vk' \rangle = (2\pi)^2
\delta(\vk-\vk')$ and $\langle \vr | \vk \rangle = e^{i\vk\vr}$, so
that the completeness relations read $\int d\vr | \vr \rangle \langle
\vr | = \unit$ and $\int \dfrac{d\vk}{(2\pi)^2} | \vk \rangle \langle
\vk | = \unit$. Using Eq.~(\ref{2.7}) together with the expression for
the free particle propagator matrix element,
\begin{equation}
\langle \vk | G_0 | \vk' \rangle = \frac{2m}{\hbar^2}
\frac{\delta(\vk-\vk')}{\kpp^2 - k^2 + i0},
\label{2.8}
\end{equation}
Correia \cite{corr} has calculated the matrix element describing a
sequence of $n$ successive collisions of the particle with scatterers
$s,r,q,\ldots,p,j$ and $i$:
\begin{equation}
\begin{split}
  &\langle \vk | T_i G_0 T_j G_0 T_p G_0 \ldots T_q G_0 T_r G_0 T_s | \vk' \rangle\\ &=
  (-1)^n \: 4i \frac{\hbar^2}{2m}e^{-i\vk\vR_i+i\vk'\vR_s}\\ &\times
  \!\!\!\!\!\!\!\!  \sum_{l_i,l_j,l_p\ldots,l_q,l_r,l_s=-\infty}^{+\infty} \left[
    \frac{J_{l_i}(ka)}{H_{l_i}^{(1)}(\kpp a)} e^{i l_i
      (\tht_k-\tht_{ij}+\pi/2)} \right] H_{l_i-l_j}^{(1)}(\kpp R_{ij})\\
  &\phantom{++} \times \left[ \frac{J_{l_j}(\kpp a)}{H_{l_j}^{(1)}(\kpp a)}
    e^{i l_j (\tht_{ij}-\tht_{jp})} \right] H_{l_j-l_p}^{(1)}(\kpp
  R_{jp}) \ldots\\ &\phantom{++} \times \left[ \frac{J_{l_r}(\kpp
      a)}{H_{l_r}^{(1)}(\kpp a)} e^{i l_r
      (\tht_{qr}-\tht_{rs})} \right] H_{l_r-l_s}^{(1)}(\kpp R_{rs})\\
  &\phantom{+++++++}\times\left[ \frac{J_{l_s}(k'a)}{H_{l_s}^{(1)}(\kpp a)}
    e^{i l_s (\tht_{rs}-\tht_{k'}-\pi/2)} \right].
\end{split}
\label{2.9}
\end{equation}
Here $(R_{ij},\tht_{ij})$ are the polar coordinates of the scatterer
separation vectors $\vR_{ij} \equiv \vR_i-\vR_j$, where
$i,j=1,2,3,\ldots,N$.

Equation~(\ref{2.5}) together with (\ref{2.9}) allows one to calculate
the energy-domain autocorrelation amplitude for an initial quantum
state $|\phi_0 \rangle$,
\begin{equation}
\begin{split}
  \Omega(E) &\equiv \langle \phi_0 | G(E) | \phi_0 \rangle\\ &= \int \frac{d\vk}{(2\pi)^2} \int
  \frac{d\vk'}{(2\pi)^2} \langle \phi_0 | \vk \rangle \langle \vk | G(E) | \vk' \rangle \langle \vk' |
  \phi_0 \rangle.
\end{split}
\label{2.10}
\end{equation}
The time-domain autocorrelation function, Eq.~(\ref{intr-1}), is then
obtained from the energy-dependent overlap with the help of the
inverse Fourier transform given by Eq.~(\ref{2.4}), namely
\begin{equation}
C(t) = \left| \Omega(t) \right|^2,
\label{2.10.5}
\end{equation}
with the time-domain autocorrelation amplitude
\begin{equation}
\Omega(t) \equiv \langle \phi_0 | G(t) | \phi_0 \rangle
=\int_{-\infty}^{+\infty} \frac{dE}{2\pi} e^{-iEt/\hbar} \Omega(E).
\label{2.11}
\end{equation}

\subsection{Wave packet}

In order to calculate the autocorrelation function due to the
propagator given by Eqs.~(\ref{2.5}) and (\ref{2.9}) we consider a
circular wave packet defined by
\begin{equation}
\phi_0(\vr) = \frac{e^{i\vk_0\vr}}{\sqrt{\pi \sigma^2}} 
\: \Theta(\sigma-|\vr-\vR_0|).
\label{2.12}
\end{equation}
The wave packet represents a particle, with the average momentum
$\hbar\vk_0$, located at the position $\vR_0$. Here $\Theta$ is the Heaviside
step function, and the real quantity $\sigma$ has an apparent meaning of
the wave packet dispersion. Our choice of the wave packet is motivated
by the demand to facilitate complex energy plane integration in the
transformation of the autocorrelation overlap from energy to time
domain, Eq.~(\ref{2.11}). As we will see in Section III the time decay
of the wave packet autocorrelation function does not depend
significantly on the initial wave packet, as long as the latter is
sufficiently localized in the coordinate and momentum space.

Let us now fix the system of coordinates by imposing $\vR_0=0$, so
that the particle is initially located at the origin. The circular
wave packet does not overlap with any of the disk scatterers if the
following $N$ conditions are satisfied:
\begin{equation}
R_j > a + \sigma, \;\;\;\;\; \mathrm{for} \;\;\; j=1,2,3,\ldots,N.
\label{2.13}
\end{equation}
The momentum representation of the wave packet is given by
\begin{equation}
\begin{split}
  &\phi_0(\vk,\vk_0) =2\sqrt{\pi} \: \frac{J_1(|\vk-\vk_0|
    \sigma)}{|\vk-\vk_0|}\\ &\equiv \phi_0(k,k_0;\tht_k-\tht_{k_0})
  =\sum_{l=-\infty}^{+\infty}\chi_l(k,k_0)e^{il(\tht_k-\tht_{k_0})},
\end{split}
\label{2.14}
\end{equation}
with
\begin{equation}
\chi_l(k,k_0) = 2\sqrt{\pi} \: \frac{k J_{l+1}(k\sigma) J_l(k_0\sigma)
  - k_0 J_l(k\sigma)J_{l+1}(k_0\sigma)}{k^2-k_0^2}.
\label{2.15}
\end{equation}
Here $(k_0,\tht_{k_0})$ are the polar coordinates of the wave vector
$\vk_0$. Appendix A provides the derivation of the above expansion.

We are now in a position to calculate the part of the wave packet
autocorrelation overlap due to a sequence of $n$ successive collisions
of the particle initially at $\vR_0$ with scatterers
$s,r,q,\ldots,p,j$ and $i$. Performing the integration over the
$\vk$-space we obtain
\begin{equation}
\begin{split}
  &\langle \phi_0 | G_0 T_i G_0 T_j G_0 T_p G_0 \ldots T_q G_0 T_r G_0 T_s G_0 | \phi_0
  \rangle\\ &= (-1)^n \frac{1}{4i}\frac{2m}{\hbar^2}\\ &\times \!\!\!
  \sum_{l,l_i,\ldots,l_s,l'=-\infty}^{+\infty} \left[\chi_l(\kpp,k_0)
    e^{il(\tht_{0i}-\tht_{k_0}-\pi/2)} \right]^* H_{l-l_i}^{(1)}(\kpp
  R_{0i})\\&\phantom{++}\times \left[ \frac{J_{l_i}(\kpp
      a)}{H_{l_i}^{(1)}(\kpp a)} e^{i l_i (\tht_{0i}-\tht_{ij})}
  \right] H_{l_i-l_j}^{(1)}(\kpp R_{ij})\\ &\phantom{++}\times \left[
    \frac{J_{l_j}(\kpp a)}{H_{l_j}^{(1)}(\kpp a)} e^{i l_j
      (\tht_{ij}-\tht_{jp})} \right] H_{l_j-l_p}^{(1)}(\kpp R_{jp}) \ldots\\
  &\phantom{++}\times \left[ \frac{J_{l_s}(\kpp a)}{H_{l_s}^{(1)}(\kpp a)}
    e^{i l_s (\tht_{rs}-\tht_{s0})} \right] H_{l_s-l'}^{(1)}(\kpp
  R_{s0})\\ &\phantom{++++++}\times \left[ \chi_{l'}(\kpp,k_0)
    e^{il'(\tht_{s0}-\tht_{k_0}-\pi/2)} \right].
\end{split}
\label{2.16}
\end{equation}
The polar coordinates of the disk separation vectors $\vR_{0i} \equiv
\vR_0-\vR_i$ and $\vR_{s0} \equiv \vR_s-\vR_0$ are given by
$(R_{0i},\tht_{0i})$ and $(R_{s0},\tht_{s0})$ respectively; the
asterix denotes the complex conjugate.

\bigskip

\subsection{Diffraction regime approximation}

The expression for the overlap in Eq.~(\ref{2.16}) is exact. We will
now derive an approximation of this overlap for the case of a dilute
scattering system. In the dilute hard-disk billiard, $R_{ij} \gg a$,
so that the
high argument approximation of the Hankel functions by exponentials
can be used to greatly simplify the analysis of the autocorrelation
function.

\begin{figure}[h]
  \centerline{\epsfig{figure=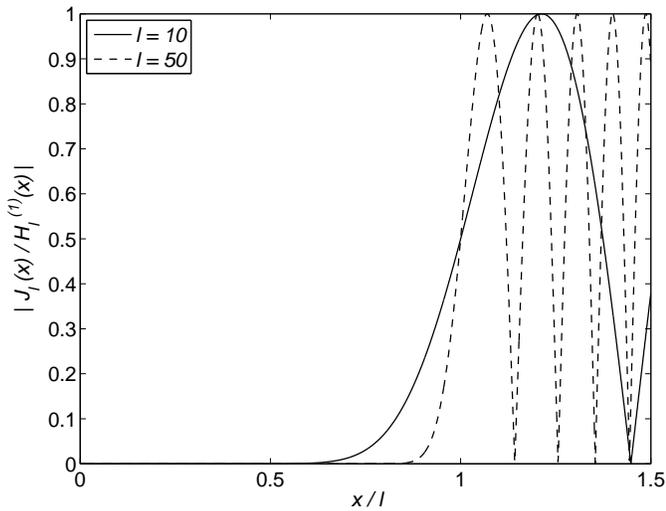,width=3.5in}}
\caption{The absolute value of the ratio $J_l(x)/H_l^{(1)}(x)$ as a
  function of $x/l$ for two cases: $l=10$ and $l=50$. The ratio
  significantly differs from zero only for $x \agt l$.}
\label{fig-0}
\end{figure}

Let us start with determining the angular momentum states dominating
the autocorrelation function for a given value of $\kpp$.  The ratios
inside the brackets in Eq.~(\ref{2.16}), $J_l(x)/H_l^{(1)}(x)$, are
small for $l > x$, since $J_l(x)$ has its first maximum at $x\sim l$.
(Figure~\ref{fig-0} illustrates the latter argument for cases of
$l=10$ and $l=50$.) Consequently, the main contribution to the
multiple sum in Eq.~(\ref{2.16}) comes from terms with $l_i,l_j,\ldots,l_s$
running from $-[\kpp a]$ to $+[\kpp a]$, and $l,l'$ running from
$-[\kpp\sigma]$ to $+[\kpp\sigma]$, where the square brackets denote the integer
part. At the same time, the large argument approximation of the Hankel
function \cite{arfk,gradst},
\begin{equation}
H_l^{(1)}(x)\approx \sqrt{\frac{2}{\pi ix}} \exp \left[ i\left(
    x-\frac{\pi l}{2} \right) \right],
\label{2.17}
\end{equation}
holds for
\begin{equation}
x \gg \frac{1}{2} \left( l^2-\frac{1}{4} \right).
\label{2.18}
\end{equation}
Therefore, if $\kpp R_{ij} \gg (2\kpp a)^2/2$, and $\kpp R_{0i}, \kpp
R_{s0} \gg (\kpp a + \kpp\sigma)^2/2$ we can use this approximation in
Eq.~(\ref{2.16}) to get
\begin{equation}
\begin{split}
  \langle \phi_0 &| G_0 T_i G_0 T_j G_0 \ldots T_r G_0 T_s G_0 | \phi_0 \rangle \approx
  -\frac{m}{\hbar^2} \left(
    \frac{i}{2\pi\kpp}\right)^{1/2}\\
  &\times\phi_0^*(\kpp,k_0;\tht_{0i}-\tht_{k_0}) \frac{e^{i\kpp
      R_{0i}}}{\sqrt{R_{0i}}}
  f_\kpp(\tht_{0i}-\tht_{ij})\frac{e^{i\kpp R_{ij}}}{\sqrt{R_{ij}}}
  \ldots\\ &\times f_\kpp(\tht_{rs}-\tht_{s0}) \frac{e^{i\kpp
      R_{s0}}}{\sqrt{R_{s0}}} \phi_0(\kpp,k_0;\tht_{s0}-\tht_{k_0}),
\end{split}
\label{2.19}
\end{equation}
where
\begin{equation}
f_\kpp(\tht) = - \left( \frac{2}{\pi i\kpp} \right)^{1/2}
\sum_{l=-\infty}^{+\infty} \frac{J_l(\kpp a)}{H_l^{(1)}(\kpp
a)}e^{il\tht}
\label{2.20}
\end{equation}
is the scattering amplitude \cite{joach} describing scattering of a
quantum particle of energy $E=\hbar^2\kpp^2/2m$ from a hard disk of
radius $a$ at an angle $\tht$.

The approximation given by Eq.~(\ref{2.19}) is only valid for energies
satisfying the conditions (see Eq.~(\ref{2.18}) and the discussion
right below it)
\begin{equation}
\kpp \ll \frac{R_{ij}}{2a^2} \: , \frac{2R_{0i}}{(a+\sigma)^2}
\;\;\;\;\; \mathrm{for} \;\;\;i,j=1,2,3,\ldots,N.
\label{2.21}
\end{equation}
These conditions bear a simple physical meaning. Suppose
$R_{ij},R_{0i}\sim R$ and $\sigma\sim a$, then inequalities~(\ref{2.21}) can be
written as $R \gg 2a/\alpha$, with $\alpha = 1/\kpp a$. The latter has a meaning
of the angle of diffraction of a wave with the wave length $1/\kpp$ on
an obstacle of size $a$, so that $2a/\alpha$ represents the estimate of the
shadow depth, which is the largest distance over which the geometrical shadow
can exist. Then, the conditions~(\ref{2.21}) simply mean that the
scattering system is so dilute that the average separation between
scatterers is much greater than the shadow depth for the given
particle's energy. This is equivalent to stating that Eq.~(\ref{2.19})
is only valid in the {\it diffraction regime}, i.e. the diffraction
effects prevail over the geometrical shadow, so that no disk scatterer
can be screened from the particle by other disks. In the case of a
sufficiently dilute scattering system, $R\gg a$, the
inequalities~(\ref{2.21}) are satisfied for a significant range of
energies, so that Eq.~(\ref{2.19}) provides a good approximation of
the autocorrelation function for wave packets confined to this energy
range.

Equation~(\ref{2.19}) has an apparent structure. The initial wave
packet $|\phi_0 \rangle$ is propagated through a sequence of free flight and
collision events, each of which contributes by a corresponding product
term to the expression for the autocorrelation overlap. Indeed, the
latter is multiplied by $e^{i\kpp R}/\sqrt{R}$ each time the particle
experiences a free flight through some length $R$, and by
$f_\kpp(\tht)$ each time the particle scatters off a disk at an angle
$\tht$. The same wave function construction algorithm was earlier used
in reference \cite{gasp-II} for semiclassical quantization or the
three-disk scattering problem. Nevertheless, it is important to note
that, at least for the autocorrelation function calculation, this
method fails in the true semiclassical limit, $\kpp \to \infty$, due to
violation of the conditions given by (\ref{2.21}), and is legitimate
only in the diffraction regime approximation defined by (\ref{2.21}).

A closer examination of Eq.~(\ref{2.19}) shows that only such initial
wave packets that represent particles moving in the vicinity of
system's classical periodic orbits can exhibit a significantly
non-zero autocorrelation function for times corresponding to a number
of collision events of the counterpart classical particle. To see
that, suppose that the wave packet $|\phi_0 \rangle$ given by
Eq.~(\ref{2.14}) is well localized in momentum space, i.e. the
particle's de Broglie wave length $\la\equiv 1/k_0 \ll \sigma$ . Then,
the overlap given by Eq.~(\ref{2.19}) is negligible unless $\tht_{0i}
\approx \tht_{s0} \approx \tht_{k_0}$. Indeed, $\phi_0(k,k_0;\tht)$,
if considered as a function of the angle $\tht$ between the wave
vectors $\vk$ and $\vk_0$, is sharply peaked at $\tht=0$, and rapidly
vanishes as $\vk$ turns away from $\vk_0$. This means that the
autocorrelation overlap $\langle \phi_0 | G | \phi_0 \rangle$ gets a
significant scattering contribution in addition to its free-streaming part,
$\langle \phi_0 | G_0 | \phi_0 \rangle$, (which is negligible for
dilute systems) only if the wave packet is initially located on and
moves along a line connecting centers of any two disks in the
scattering system. This happens because the reflection wave produced
by a disk at the last scattering interferes destructively with the
initial wave unless the two waves have their wave vectors pointing
almost in the same direction. Hence, one can expect substantially
non-zero values of the autocorrelation function only for such initial
wave packets that represent classical particles moving along lines
connecting scatterer centers, and therefore traveling in the vicinity
of unstable periodic orbits of the hard-disk scattering system.

We will now use the propagator given by Eqs.~(\ref{2.5}) and
(\ref{2.19}) to calculate the autocorrelation function for the
circular wave packet defined in Eq.~(\ref{2.14}) in three different
scattering systems: two-disk, three-disk equilateral and three-disk
isosceles billiards. The autocorrelation overlap in the energy domain
can be written as
\begin{equation}
\Omega(E) = \langle \phi_0 | G | \phi_0 \rangle = \Omega_0(E) +
\Omega_\mathrm{S}(E),
\label{2.22}
\end{equation}
where $\Omega_0(E)=\langle \phi_0 | G_0| \phi_0 \rangle$ is the free flight
part of the overlap, and the scattering part, $\Omega_\mathrm{S}(E)$, is
determined by all possible collision events:
\begin{equation}
\begin{split}
  \Omega_\mathrm{S}(E) &= \sum_j \langle \phi_0 | G_0 T_j G_0 | \phi_0 \rangle\\
  &+ \sum_j \sum_{k \neq j} \langle \phi_0 | G_0 T_j G_0 T_k G_0 |\phi_0 \rangle\\ &+ \sum_j \sum_{k \neq
    j} \sum_{l \neq k} \langle \phi_0 | G_0 T_j G_0 T_k G_0 T_l G_0 | \phi_0 \rangle + \ldots.
\end{split}
\label{2.23}
\end{equation}
The series in Eq.~(\ref{2.23}) can be summed explicitly for the three
above-mentioned billiard systems using the diffraction regime
approximation, Eq.~(\ref{2.21}), together with the assumption of that
the initital wave packet has sufficiently high energy,  as will be described below.

\subsection{Two-disk billiard}

\begin{figure}[h]
  \centerline{\epsfig{figure=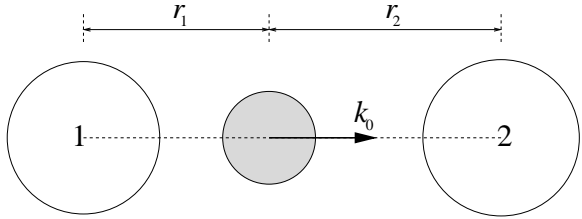,width=3.0in}}
  \caption{Two-disk billiard. The circular wave packet is initially
    located on the classical periodic orbit distance $r_1$ away from
    the center of disk ``1'', and distance $r_2$ away from the center
    of disk ``2'', with $r_1+r_2=R$.}
\label{fig-1}
\end{figure}

The simplest of all hard disk systems is the two-disk billiard, which
consists of two hard disks, ``1'' and ``2'', of radius $a$ with the
center-to-center separation $R$, see fig.~\ref{fig-1}. Following the
discussion above we place the initial wave packet on the line
connecting the disk centers with its wave vector $\vk_0$ pointing
along this line in order for the sum in Eq.~(\ref{2.23}) to have
significant, non-vanishing terms. For the initial condition shown in
fig.~\ref{fig-1} these non-vanishing terms are
\begin{equation}
\begin{split}
  &\langle \phi_0 | G_0 T_1 G_0 T_2 G_0 \ldots T_1 G_0 T_2 G_0 | \phi_0 \rangle\\ &=
  -\frac{m}{\hbar^2} \left( \frac{i}{2\pi\kpp}\right)^{1/2} \left|
    \phi_0(\kpp,k_0;0) \right|^2\\ &\times \frac{e^{i\kpp r_1}}{\sqrt{r_1}}
  f_\kpp(\pi) \frac{e^{i\kpp R}}{\sqrt{R}}f_\kpp(\pi) \frac{e^{i\kpp
      R}}{\sqrt{R}} \ldots f_\kpp(\pi) \frac{e^{i\kpp R}}{\sqrt{R}} f_\kpp(\pi)
  \frac{e^{i\kpp r_2}}{\sqrt{r_2}},
\end{split}
\label{2.24}
\end{equation}
where the diffraction regime approximation, Eq.~(\ref{2.19}), has been
used. Here, $r_1$ and $r_2$ are the distances separating the center of
the wave packet and the centers of disks ``1'' and ``2'' respectively;
$r_1 + r_2 = R$. Substituting Eq.~(\ref{2.24}) into Eq.~(\ref{2.23}),
while neglecting all other scattering sequences, one obtains a
geometric series that sums to
\begin{equation}
\Omega_\mathrm{S}(E) = \frac{m}{\hbar^2} \left( \frac{iR}{2\pi\kpp r_1 r_2}
\right)^{1/2} \frac{\left| \phi_0(\kpp,k_0;0) \right|^2}{1-\left(
f_\kpp(\pi) \dfrac{e^{i\kpp R}}{\sqrt{R}} \right)^{-2}}.
\label{2.25}
\end{equation}
Equation~(\ref{2.21}) along with the assumption $r_1\sim r_2\sim R/2$ shows
that the validity of the result predicted by Eq.~(\ref{2.25}) is
limited to the energies satisfying $\kpp\ll R/2a^2,R/(a+\sigma)^2$. Now, if
the wave packet is sufficiently localized in the momentum space, i.e.
the de Broglie wave length
\begin{equation}
\la \equiv 1/k_0 \ll \sigma,
\label{2.25.5}
\end{equation}
and if $k_0\ll R/2a^2,R/(a+\sigma)^2$, then Eq.~(\ref{2.25}) can be
used to calculate the time domain autocorrelation function for the
wave packet.

Finally we make an assumption that the quantum particle is highly
energetic, so that its de Broglie wave length is much smaller than the
scatterer size, i.e. $\la \ll a$. This approximation allows us to use
the semiclassical (WKB) expression for the hard-disk scattering
amplitude, namely
\begin{equation}
f_\kpp(\tht) = -\sqrt{\frac{a}{2}|\sin(\tht/2)|} \: e^{-2i\kpp
a|\sin(\tht/2)|}.
\label{2.26}
\end{equation}
Equation~(\ref{2.26}) is known to be a good approximation of the exact
scattering amplitude for sufficiently large scattering angles
\cite{nuss}. Then, combining the diffraction approximation,
Eq.~(\ref{2.21}), with the short de Broglie wavelength approximation,
$\la \ll a$, we arrive at the following
condition:
\begin{equation}
\frac{2a^2}{R},\dfrac{(a+\sigma)^2}{R} \ll \la \ll a.
\label{2.27}
\end{equation}
We refer to this inequality as to the condition of the {\it
high-energy diffraction regime}. In the rest of this paper we assume
that both the momentum space localization condition,
Eq.~(\ref{2.25.5}), and the high-energy diffraction regime condition,
Eq.~(\ref{2.27}), are satisfied.

The calculation of the autocorrelation function in the time domain,
$C(t)$, is carried out in accordance with Eqs.~(\ref{2.10.5}),
(\ref{2.11}), (\ref{2.22}) and (\ref{2.25}). The scattering part of
the time-domain autocorrelation amplitude reads
\begin{equation}
\Omega_\mathrm{S}(t) = \frac{i}{2\pi} \frac{\hbar^2}{m} \int_\Gamma d\kpp
\kpp \exp \left( -i\frac{\hbar t}{2m}\kpp^2 \right) \Omega_\mathrm{S}(E),
\label{2.28}
\end{equation}
where $\Omega_\mathrm{S}(E)$ is given by Eq.~(\ref{2.25}), and contour
$\Gamma$ in the complex $\kpp$-plane follows the imaginary axis from
$-i\infty$ to $0$, then turns at the right angle, and proceeds to
$+\infty$ along the real axis.

Careful analytical calculation of a range of times, for which
Eq.~(\ref{2.28}), with $\Omega_\mathrm{S}(E)$ given by Eq.~(\ref{2.25}),
yields accurate predictions, is a formidable problem. Nevertheless, we
can use simple physical arguments to roughly estimate this time range.
The method that we used to calculate the energy-dependent scattering
part of the autocorrelation function, $\Omega_\mathrm{S}(E)$, relies on the
assumption of the wave packet diffraction.  The wave packet must
explore the scattering system for the diffraction effects to take
place.  An estimate of the time needed for the particle to reach the
first scatterer is $t_E \approx r_2/v \sim R/v$, where $r_2$ is the distance
between the initial location of the particle and the first scatterer
it collides with, see fig.~\ref{fig-1}, and $v=\hbar k_0/m$ is the average
velocity of the wave packet. Time $t_E$ also gives an estimate of the
Ehrenfest time for the system: for times shorter than $t_E$ the wave
packet evolution is dominated by the free particle Hamiltonian, and
therefore is classical-like, while particle's propagation is
diffractive and substantially non-classical for times beyond $t_E$. To
estimate the upper bound of the applicability time-range we note that
the WKB expression for the scattering amplitude, Eq.~(\ref{2.26}),
breaks down at low energies $E=\hbar^2 \kpp^2/2m$ with $\kpp \sim 1/a$. The
momentum corresponding to these energies is $\hbar \kpp \sim (\la/a) \hbar k_0$,
and the corresponding velocity is $v_\kpp \sim (\la/a) v$. The
contribution of these low energy modes of the particle to the
autocorrelation function become significant after the long wavelength
part of the wave packet explores the scattering system, {\it i.e.}
after times $t_{\max} \sim R/v_\kpp \sim (a/\la) R/v$ corresponding to
$a/\la$ particle-disk collisions. Since $a \gg \la$ this number of
collisions can be quite large. Summarizing, we find that the
applicability time-range of our analysis of the time-dependent
autocorrelation function is estimated by
\begin{equation}
1 \alt vt/R \alt a/\la.
\label{2.29}
\end{equation}

In order to evaluate the integral in Eq.~(\ref{2.28}) one can show
that for $t>0$ the contour can be closed along the infinite
quarter-circle $\kpp = \kpp_\infty e^{i\gamma}$, with $\kpp_\infty
\to +\infty$, and the angle $\gamma$ decreasing from $0$ to
$-\pi/2$ \footnote{We have chosen an initial wave packet that allows us
to close the integration contour. This would not be possible
for a Gaussian initial wave packet, and one would need to employ
different methods to calculate the integral in Eq.~(\ref{2.28}).}, so
that the value of the integral is determined by poles of
$\Omega_\mathrm{S}(E)$ in the fourth quadrant of the complex $\kpp$
plane:
\begin{equation}
\kpp_n = \frac{\pi n}{R} - \frac{i}{2R} \ln\frac{2R}{a},
\;\;\;\;\;\;\; \mathrm{for} \;\;\; n=1,2,3,\ldots.
\label{2.30}
\end{equation}
The semiclassical approximation for the scattering amplitude,
Eq.~(\ref{2.26}), was used to find zeros of the denominator of
$\Omega_\mathrm{S}(E)$, so that Eq.~(\ref{2.30}) correctly locates the
poles right below the region on real $\kpp$ axis where
$\phi_0(\kpp,k_0;0)$ is localized.  Then, calculating the residues
corresponding to the poles, we get
\begin{equation}
\begin{split}
  &\Omega_\mathrm{S}(t) \approx \frac{1}{2}\left( 2\pi i R r_1 r_2 \right)^{-1/2}\\
  &\times \sum_{n \, \sim \, n_0-[R/\sigma]}^{n_0+[R/\sigma]} \sqrt{\kpp_n}\left|
    \phi_0(\kpp_n,k_0;0) \right|^2 \exp\left( -i\frac{\hbar t}{2m}\kpp_n^2
  \right),
\end{split}
\label{2.31}
\end{equation}
where $\pi n_0/R=k_0$, and the square brackets in the summation limits
represent the integer part. Equation~(\ref{2.31}) is expected to hold
within the time range given by (\ref{2.29}).

The free streaming part of the autocorrelation function overlap,
$\Omega_0(t)=\langle \phi_0 | G_0(t) | \phi_0 \rangle$, is calculated
explicitly for the free particle propagator. It can be shown
that its contribution to the full autocorrelation
function $C(t)$ is negligible for dilute billiard systems
\footnote{The decay of the free streaming part of the autocorrelation
  overlap, $\Omega_0(t)$, results from the spatial separation of the
  wave packets at times $0$ and $t$, and is predominantly
  Gaussian. Reference \cite{disser} estimates the magnitude of this
  overlap to be thousands orders of magnitude smaller than the one of
  the scattering part of the autocorrelation overlap,
  $\Omega_\mathrm{S}(t)$, for the set of parameters used to illustrate
  the $C(t)$ decay throughout this paper, i.e. $a=\sigma=1$, $R=10^4$
  and $\la=10^{-2}$.}. Therefore, the autocorrelation function for
  long times is entirely determined by the scattering events, so that
  $C(t) \approx |\Omega_\mathrm{S}(t)|^2$.

The main features of the time-domain autocorrelation function $C(t)$
can be deduced by considering only a small number of poles with
$n=n_0+\tilde{n}$, where $\tilde{n}$ is sufficiently small, so that
the pre-exponential function in Eq.~(\ref{2.31}) stays approximately
constant. The contribution due to these poles is
\begin{equation}
\begin{split}
\Omega_\mathrm{S}(t) &\sim \sum_{\tilde{n}} \exp\left[ -i\frac{\hbar t}{2m}
\left( k_0+\frac{\pi\tilde{n}}{R}-\frac{i}{2R}\ln\frac{2R}{a}
\right)^2 \right]\\ &\sim e^{-iE_0 t/\hbar} \exp\left(-\frac{1}{2}
\lambda^{(2)} t \right) \sum_{\tilde{n}}
e^{-i\frac{vt}{R}\pi\tilde{n}},
\end{split}
\label{2.32}
\end{equation}
where $E_0=\hbar^2 k_0^2/2m$ is the average energy of the wave packet,
$v=\hbar k_0/m$ is its average velocity, and
\begin{equation}
\lambda^{(2)}=\dfrac{v}{R}\ln \dfrac{2R}{a}
\label{2.33}
\end{equation}
is the classical Lyapunov exponent of the two-disk periodic orbit in
the limit $R\gg a$, e.g. see reference \cite{gasp-book}. Equation
(\ref{2.32}) shows that (i) the envelope of the scattering part of the
autocorrelation function decays exponentially with time, $C(t) \sim
e^{-\lambda^{(2)} t}$, with the decay rate given by the classical Lyapunov
exponent of the system, and (ii) strong interference peaks occur in
$C(t)$ at times $t$ multiple to the period of the classical periodic
orbit, i.e. when $vt/R$ is an even integer. The peaks of the
autocorrelation function correspond to partial reconstruction of the
wave packet at times at which the counterpart classical particle
returns to its initial point in the phase space.

\begin{figure}[h]
  \centerline{\epsfig{figure=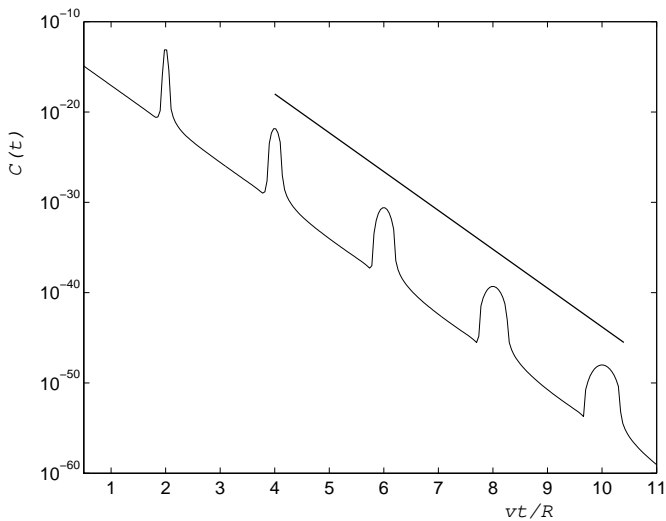,width=3.5in}}
  \caption{Wave packet autocorrelation function $C(t)$ vs. $vt/R$ for
    the two-disk billiard. Parameters of the system are as follows:
    $a=\sigma=1$, $R=10^4$, $r_1=r_2=R/2$ and $\la=10^{-2}$. The straight
    line shows exponential decay with the rate given by the classical
    two-disk Lyapunov exponent $\lambda^{(2)}$. The decay is shown for times
    $t$ greater than the Ehrenfest time $t_E\approx R/2v$.}
\label{fig-2}
\end{figure}

Figure~\ref{fig-2} shows the autocorrelation function, $C(t)$,
computed using Eq.~(\ref{2.31}), with the wave packet given by
Eq.~(\ref{2.14}), for the two-disk billiard with the following
parameters: $a=1$ and $R=10^4$. The circular wave packet of size
$\sigma=1$ is initially placed in the middle between the disks,
$r_1=r_2=R/2$, and its de Broglie wave length
$\la=10^{-2}$. Conditions~(\ref{2.25.5}) and (\ref{2.27}) are
satisfied by this system. The summation in Eq.~(\ref{2.31}) includes
20,001 poles. The solid line shows $e^{-\lambda^{(2)} t}$ decay. The
figure shows the decay for times $t$ greater than the Ehrenfest time
$t_E\approx R/2v$.

As mentioned above, fig.~\ref{fig-2} exhibits the wave packet
reconstruction (revival) peaks together with the exponential decay of
the autocorrelation function envelope. Another distinct feature of the
decay is the broadening of the peaks in the course of time. These
phenomena have a simple physical explanation. The direction along the
two-disk periodic orbit is neutrally stable, i.e. a perturbation of
the initial phase-space point (of a classical particle on the periodic
orbit) along this direction growths at most linearly with time. On the
other hand, the direction perpendicular to the periodic orbit is
exponentially unstable, with the Lyapunov exponent $\lambda^{(2)}$
playing the role of the classical instability rate. Therefore, the
wave packet probability density dies out linearly with time in the
direction along the periodic orbit, whereas it decreases exponentially
in the perpendicular direction \footnote{The situation is similar to
the one observed in the short time spreading of small Gaussian wave
packets in the Lorentz gas systems, e.g. see \cite{us}.}. The
spreading of the wave packet in the direction along the periodic orbit
(and therefore along its average velocity) yields prolongation of time
intervals during which $|\phi_t\rangle$ significantly overlaps with
the initial state $|\phi_0\rangle$. This prolongation amounts to
broadening of the revival peaks of the autocorrelation function. On
the other hand, the wave packet spreading in the unstable direction is
responsible for the exponential decay of the autocorrelation function
envelope.

\subsection{Three-disk systems}

We will now calculate the autocorrelation function decay for different
scattering systems, namely three-disk billiards, in which a particle
moves in two dimensional space with three fixed hard-disk scatterers.
Here we restrict ourselves only to such three-disk billiards for which
the centers of the disks constitute vertices of an equilateral or of
an isosceles triangle. The corresponding scattering systems are then
referred to as the three-disk equilateral and isosceles billiards
respectively. The discussion of the generic three-disk billiard, in
which all three sides of the triangle are different, is left for the
Section~III.

\subsubsection{Three-disk equilateral billiard}

\begin{figure}[h]
\centerline{\epsfig{figure=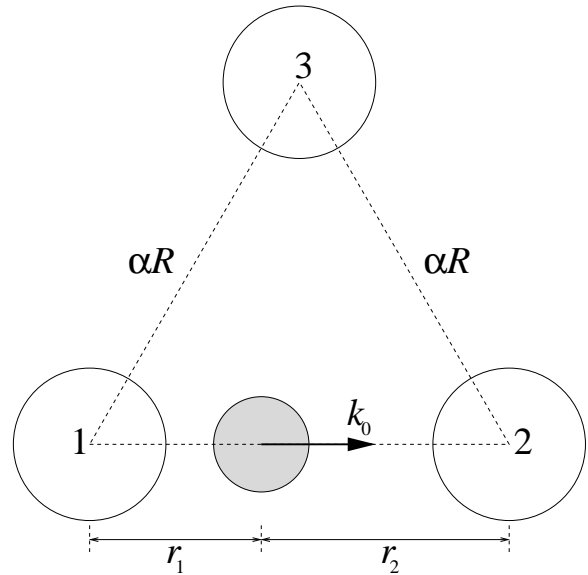,width=3.0in}}
\caption{Three-disk isosceles billiard. The circular wave packet is
  initially located distance $r_1$ away from disk ``1'', and distance
  $r_2$ away from disk ``2'', with $r_1+r_2=R$. Disk ``3'' is distance
  $\alpha R$ away from disks ``1'' and ``2''. The ``equilateral''
  billiard case corresponds to $\alpha=1$.}
\label{fig-3}
\end{figure}

We consider a geometrically open billiard consisting of three hard
disks of radius $a$ centered at the vertices of an isosceles triangle
with one side of length $R$ and two other sides of length $\alpha R$,
see fig.~\ref{fig-3}. Here we focus on the case $\alpha=1$, which
allows complete analytical treatment. The following section is devoted
to the case $\alpha\neq 1$, for which a substantial understanding can
be achieved in the limit $\alpha\gg 1$.

As in the case of the two-disk billiard we initially put the wave
packet of size $\sigma$ on the line connecting the center of two
disks, labeled by ``1'' and ``2'', with center-to-center separation
$R$, see fig.~\ref{fig-3}. Distances $r_1$ and $r_2$ separating the
wave packet center and the centers of disks ``1'' and ``2''
respectively satisfy the apparent condition: $r_1+r_2=R$. The average
wave vector of the wave packet $\vk_0$ is pointing along the line
connecting the centers of disks ``1'' and ``2'' as shown in
fig.~\ref{fig-3}.

Our calculation of the wave packet autocorrelation function employs
the transition matrix method first applied by Gaspard and Rice
\cite{gasp-II} for calculation of scattering resonances in three-disk
scattering systems. Transition matrices, and related to them monodromy
matrices, have also been used in analysis of different systems by
Bogomolny \cite{bogom}, and Agam and Fishman \cite{fish}. We fist
spell out the multiple collision expansion given by Eq.~(\ref{2.23})
for the case of a three-disk billiard:
\begin{equation}
\Omega_\mathrm{S}(E) = \sum_{n=2}^\infty \: \sum_{\mathrm{path}(n)}\langle
\phi_0 | G_0 T_1 G_0 T_i G_0 T_j \ldots G_0 T_2G_0 | \phi_0\rangle,
\label{2.34}
\end{equation}
where the second sum runs over all possible paths consisting of $n$
binary collision events with the first collision taking place at disk
``2'' and the $n^{\mathrm{th}}$ collision at disk ``1''. Every term in
this double sum is evaluated using the diffraction regime
approximation, Eq.~(\ref{2.19}). Following \cite{gasp-II} we construct
a $6\times 6$ matrix $\vQ$ describing a transition in 6-dimensional
space composed of directions $(1\!\to\!  2)$,
$(1\!\to\!  3)$, $(2\!\to\! 1)$, $(2\!\to\!
3)$, $(3\!\to\! 1)$ and $(3\!\to\! 2)$, due to a
single scattering event,
\begin{equation}
\vQ = 
\begin{array}{cc}
  ^{1\cdot 2} \;\; ^{1\cdot 3} \;\; ^{2\cdot 1} \;\; ^{2\cdot
    3} \;\; ^{3\cdot 1} \;\; ^{3\cdot 2} & \\ \left(
\begin{array}{cccccc}
0 & 0 & X & W & 0 & 0\\ 0 & 0 & 0 & 0 & X & W\\ X & W & 0 & 0 & 0 &
0\\ 0 & 0 & 0 & 0 & W & X\\ W & X & 0 & 0 & 0 & 0\\ 0 & 0 & W & X & 0
& 0
\end{array}\right) &
\begin{array}{c}
\!\! ^{1\cdot 2}\\
\!\! ^{1\cdot 3}\\
\!\! ^{2\cdot 1}\\
\!\! ^{2\cdot 3}\\
\!\! ^{3\cdot 1}\\
\!\! ^{3\cdot 2}
\end{array}
\end{array}
\label{2.35}
\end{equation}
where
\begin{equation}
X=f_\kpp(\pi) \frac{e^{i\kpp R}}{\sqrt{R}} \;\;\;\;\; \mathrm{and}
\;\;\;\;\; W=f_\kpp(2\pi/3) \frac{e^{i\kpp R}}{\sqrt{R}}.
\label{2.36}
\end{equation}
Here $\pi$ and $2\pi/3$ are the turning angles for a classical
particle bouncing among three disk of radius $a$ placed in the
vertices of an equilateral triangle with side $R \gg a$. The second
sum in Eq.~(\ref{2.34}) is then given by the one-one element of the
matrix $\vQ^n$:
\begin{equation}
\begin{split}
  &\sum_{\mathrm{path}(n)} \langle \phi_0 | G_0 T_1 G_0 T_i G_0 T_j \ldots G_0 T_2G_0 |
  \phi_0 \rangle\\ &\phantom{++}= -\frac{m}{\hbar^2}\left(\frac{iR}{2\pi\kpp r_1
      r_2}\right)^{1/2} \left| \phi_0(\kpp,k_0;0)\right|^2\left( \vQ^n
  \right)_{1,1}.
\end{split}
\label{2.37}
\end{equation}
Substituting Eq.~(\ref{2.37}) into Eq.~(\ref{2.34}), and taking
advantage of the equality $\sum_{n=2}^\infty \vQ^n = \vQ^2
\left(\unit-\vQ \right)^{-1}$, we obtain
\begin{equation}
\begin{split}
  &\Omega_\mathrm{S}(E) = \frac{m}{\hbar^2}\left( \frac{iR}{2\pi\kpp r_1 r_2}
  \right)^{1/2} \left|\phi_0(\kpp,k_0;0)\right|^2\\ &\times \left( 1 -
    \frac{1/6}{1-W+X} - \frac{1/6}{1-W-X} \right.\\
  &\phantom{+++++++++}\left. - \frac{(2+W)/3}{1+W+W^2-X^2} \right).
\end{split}
\label{2.38}
\end{equation}

As in the two-disk billiard case the poles of $\Omega_\mathrm{S}(E)$
located in the fourth quadrant of the complex $\kpp$ plane determine
the time-domain autocorrelation function. Following \cite{gasp-II} we
define $\xi \equiv -\sqrt{a/R} \: e^{i\kpp R}$ to find that in the
limit $R\gg a$ the poles of $S(E)$ are given by
\begin{equation}
\kpp_{n,j} = \frac{2\pi n + \pi + \arg\xi_j}{R} - \frac{i}{2R}
\ln\frac{R|\xi_j|^2}{a},
\label{2.39}
\end{equation}
where $j=1,2,3,4$, and
\begin{equation}
\begin{split}
  &\xi_1=\frac{1}{\left(1/2\right)^{1/2}+\left(3^{1/2}/4\right)^{1/2}},\\
  &\xi_2 = \frac{\left(
      4^{3/2}-3^{3/2}\right)^{1/2}-3^{1/4}}{2\left[1-\left(3/4\right)^{1/2}\right]}
  \: e^{i\pi},\\ &\xi_3 = \frac{\left(
      4^{3/2}-3^{3/2}\right)^{1/2}+3^{1/4}}{2\left[1-\left(3/4\right)^{1/2}\right]},\\
  & \xi_4 =
  \frac{1}{\left(1/2\right)^{1/2}-\left(3^{1/2}/4\right)^{1/2}} \:
  e^{i\pi}.
\end{split}
\label{2.40}
\end{equation}
Equations~(\ref{2.39}) and (\ref{2.40}), which locate the scattering
resonances of the three-disk system, were originally obtained by
Gaspard and Rice \cite{gasp-II}. It is interesting to note that
$\kpp_{n2}$ and $\kpp_{n3}$, being the simple poles of the
autocorrelation amplitude $\Omega_\mathrm{S}(E)$, appear as double poles in
the three-disk scattering matrix \cite{gasp-II}. The three-disk
scattering resonances have been also calculated by Cvitanovi\'c and
Eckhardt \cite{cvit} by means of the fundamental cycle expansion
technique \cite{cvit-2}.

The time-domain scattering part of the autocorrelation function,
$\Omega_\mathrm{S}(t)$, is determined following the procedure used for the
analysis of the two-disk billiard system. As before we only consider
the poles lying under the region on the real $\kpp$-axis on which the
wave packet is mainly concentrated, i.e. $n\in (n_0-[R/2\sigma], \,
n_0+[R/2\sigma])$, with $2\pi n_0/R=k_0$ and the square brackets
denoting the integer part. Calculation of residues of
$\Omega_\mathrm{S}(E)$ is straightforward. The result is given by
\begin{equation}
\begin{split}
  &\Omega_\mathrm{S}(t) \approx \frac{1}{6} \left( 2\pi i R r_1 r_2
  \right)^{-1/2}\\ &\times \sum_{n \, \sim\,n_0-[R/2\sigma]}^{n_0+[R/\sigma]} \left\{
    \sqrt{\kpp_{n1}} \left|\phi_0(\kpp_{n1},k_0;0) \right|^2 e^{-i\frac{\hbar
        t}{2m}\kpp_{n1}^2}\right.\\ &\phantom{++++++}+
  2\sqrt{\kpp_{n2}} \left| \phi_0(\kpp_{n2},k_0;0) \right|^2e^{-i\frac{\hbar
      t}{2m}\kpp_{n2}^2}\\ &\phantom{++++++}+ 2\sqrt{\kpp_{n3}}
  \left|\phi_0(\kpp_{n3},k_0;0)
  \right|^2 e^{-i\frac{\hbar t}{2m}\kpp_{n3}^2}\\
  &\left.\phantom{++++++}+ \sqrt{\kpp_{n4}} \left|
      \phi_0(\kpp_{n4},k_0;0)\right|^2 e^{-i\frac{\hbar
        t}{2m}\kpp_{n4}^2}\right\}.
\end{split}
\label{2.41}
\end{equation}
In order to predict the main features of the decay one can consider
only poles in a small vicinity of the peak of the wave function,
$n=n_0+\tilde{n}$ with $\tilde{n}\ll n_0$. This yields
\begin{equation}
\begin{split}
  \Omega_\mathrm{S}(t) &\sim \left[ \left( e^{-\tilde{\gamma}_1 t/2}
      + 2e^{-\tilde{\gamma}_3 t/2}\right) e^{-i\frac{vt}{R}\pi} \right.\\
  &\left. +2e^{-\tilde{\gamma}_2 t/2} + e^{-\tilde{\gamma}_4 t/2} \right]e^{-
    iE_0 t/\hbar} \sum_{\tilde{n}} e^{-i\frac{vt}{R} 2\pi\tilde{n}},
\end{split}
\label{2.42}
\end{equation}
where, as before, $E_0$ is the average energy of the wave packet, $v$
is its average velocity, and $\tilde{\gamma}_j =
(v/R)\ln(R|\xi_j|^2/a)$ with $j=1,2,3,4$. The slowest decay rate,
$\tilde{\gamma}_1$, playing the role of the decay rate of the
autocorrelation function envelope, can be written as
\begin{equation}
\gamma^{(3)} \equiv \tilde{\gamma}_1 = \dfrac{v}{R} \ln
\dfrac{4R}{\left[ 2^{1/2} + 3^{1/4} \right]^2 a} \approx \dfrac{v}{R}
\ln \dfrac{0.54 R}{a}.
\label{2.43}
\end{equation}

\begin{figure}[h]
  \centerline{\epsfig{figure=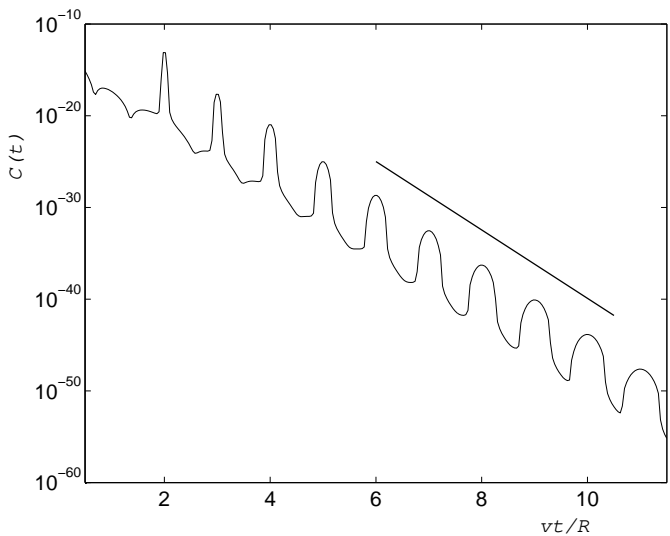,width=3.5in}}
  \caption{The autocorrelation function vs. $vt/R$ for the three-disk
    equilateral billiard. Parameters of the system are as follows:
    $a=\sigma=1$, $R=10^4$, $r_1=r_2=R/2$ and $\la=10^{-2}$. The straight
    line shows exponential decay with the rate given by $\gamma^{(3)}$. The
    decay is shown for times $t$ greater than the Ehrenfest time $t_E\approx
    R/2v$.}
\label{fig-4}
\end{figure}

Once again neglecting the free-streaming amplitude $\Omega_0(t)$ we
calculate the autocorrelation function as $C(t)\approx
|\Omega_\mathrm{S}(t)|^2$. As in the two-disk billiard case $C(t)$ exhibits
a sequence of strong wave packet revival peaks corresponding to phase
space returns of the counterpart classical particle, see
fig.~\ref{fig-4}. Taking into account that $|\xi_1|\approx 0.73$,
$|\xi_2|\approx 1.34$, $|\xi_3|\approx 11.16$ and $|\xi_4|\approx
20.38$ one can notice that the peaks occur whenever $vt/R$ is an
integer greater than one, see fig.~\ref{fig-4}. At $t=R/v$ a part of
the wave packet reflected by disk ``2'' overlaps with the initial wave
packet, but this overlap leads to destructive interference since the
wave vectors of the two waves have opposite directions.  This shows up
as the absence of the revival peak at $vt/R=1$, and appears
mathematically as partial cancellation of the expression within the
square brackets in Eq.~(\ref{2.42}). It is evident from
Eq.~(\ref{2.42}) together with the inequality $|\xi_1|,|\xi_2| \ll
|\xi_3|, |\xi_4|$ that, as pointed out by Gaspard and Rice
\cite{gasp-II}, the lines of poles corresponding to $\xi_3$ and
$\xi_4$ are screened by the other two lines of poles and have no
affect on the wave packet dynamics.  One can also show that after only
two collisions the RHS of Eq.~(\ref{2.42}) becomes totally dominated
by the first exponential term within the square brackets resulting in
the exponential decay of the autocorrelation function, $C(t) \sim
e^{-\gamma^{(3)} t}$.

Figure~\ref{fig-4} shows the decay of the time-dependent
autocorrelation function for the three-disk equilateral billiard
studied above. The parameters of the system are chosen to be identical
with once used in the case of the two-disk billiard: $a=\sigma=1$,
$R=10^4$, $r_1=r_2=R/2$ and $\la=10^{-2}$. The function $C(t)$ for a
time interval comprising 10 collision events was calculated by
computing the sum in Eq.~(\ref{2.41}) over the total of 40,006 poles.
The straight line shows $e^{-\gamma^{(3)} t}$ decay, with
$\gamma^{(3)}$ given by Eq.~(\ref{2.43}). It is interesting to note
how small the magnitude of the autocorrelation function becomes after
only a few particle-disk collisions. After the time corresponding to
ten bounces of the classical particle, the return probability drops
down to a value below $10^{-40}$ implying practical orthogonality of
the initial and final states of the quantum particle. It is the
consequence of the huge scale difference in the billiard considered
here. The phenomenon of the wave packet partial reconstruction can be
enhanced by decreasing the ratio $R/a$.

\subsubsection{Three-disk isosceles billiard}

We now address a more general three-disk scattering system, in which
the scatterers are located at the vertices of an isosceles triangle,
as shown in fig.~\ref{fig-3} with $\alpha\neq 1$. Once again the
circular wave packet is initially placed between disks ``1'' and
``2'', which have center-to-center separation $R$, while disk ``3'' is
distance $\alpha R$ away form them, see fig.~\ref{fig-3}.

The single collision transition matrix $\vQ$, which in the case of an
equilateral billiard was given by Eq.~(\ref{2.35}), now reads
\begin{equation}
\vQ = 
\begin{array}{cc}
  ^{1\cdot 2} \;\;\; ^{1\cdot 3} \;\;\; ^{2\cdot 1} \;\;\;
  ^{2\cdot 3} \;\;\; ^{3\cdot 1} \;\;\; ^{3\cdot 2} & \\ \left(
\begin{array}{cccccc}
0 & 0 & X_1 & W_1 & 0 & 0\\ 0 & 0 & 0 & 0 & X_2 & W_3\\ X_1 & W_1 & 0
& 0 & 0 & 0\\ 0 & 0 & 0 & 0 & W_3 & X_2\\ W_2 & X_2 & 0 & 0 & 0 & 0\\
0 & 0 & W_2 & X_2 & 0 & 0
\end{array}\right) &
\begin{array}{c}
\!\! ^{1\cdot 2}\\
\!\! ^{1\cdot 3}\\
\!\! ^{2\cdot 1}\\
\!\! ^{2\cdot 3}\\
\!\! ^{3\cdot 1}\\
\!\! ^{3\cdot 2}
\end{array}
\end{array}
\label{2.44}
\end{equation}
with
\begin{equation}
\begin{split}
  &X_1=f_\kpp(\pi) \frac{e^{i\kpp R}}{\sqrt{R}},
  \;\;\;\;\;\;\;\;\;\;\;\;
  X_2=f_\kpp(\pi) \frac{e^{i\kpp \alpha R}}{\sqrt{\alpha R}},\\
  &W_1=f_\kpp(\pi-\phi_1) \frac{e^{i\kpp R}}{\sqrt{R}}, \;\;\;
  W_2=f_\kpp(\pi-\phi_1) \frac{e^{i\kpp \alpha R}}{\sqrt{\alpha R}},\\
  &\phantom{+++++} W_3=f_\kpp(\pi-\phi_3) \frac{e^{i\kpp \alpha R}}{\sqrt{\alpha R}}.
\end{split}
\label{2.45}
\end{equation}
Here, $\phi_1$ and $\phi_3$ are angles of the triangle corresponding
to vertices ``1'' and ``3'' respectively; the angles are functions of
$\alpha$, and satisfy the obvious relation: $2\phi_1+\phi_3=\pi$.

Following the technique used above, we calculate the one-one element
of the matrix $\vQ (\unit-\vQ)^{-1}$ to obtain the energy dependent
autocorrelation function:
\begin{widetext}
\begin{equation}
\begin{split}
  \Omega_\mathrm{S}(E) = &-\frac{m}{\hbar^2}\left( \frac{iR}{2\pi\kpp r_1 r_2}
  \right)^{1/2} \left|\phi_0(\kpp,k_0;0)\right|^2 \: \frac{\xi^2}{\alpha^2}\\ &\times
  \frac{X_1^2 (1 - X_2^2)^2 + \left[ 2 X_1 X_2 + W_3 - X_2^2 (2 X_1
      X_2 - W_1 W_2) \right] W_1 W_2 - (X_1 X_2 - W_1 W_2)^2
    W_3^2}{\left[ 1 - X_2^2 + (X_1 X_2 - W_1 W_2) W_3 \right]^2 -
    \left[ X_1 (1 - X_2^2) +X_2 (W_1 W_2 + W_3) \right]^2}.
\end{split}
\label{2.46}
\end{equation}
\end{widetext}
As above, we introduce $\xi \equiv -\sqrt{a/R} \: e^{i\kpp R}$. The
poles of $\Omega_\mathrm{S}(E)$, i.e. zeros of the denominator in
Eq.~(\ref{2.46}), are given by solutions of the following two
polynomial equations:
\begin{equation}
\begin{split}
\left[ 2 - \left( \sqrt{2} - \sqrt{1+\frac{1}{2\alpha}} \right) \xi
  \right] \xi^{2\alpha} &= \omega_+ \left( -\frac{a}{R}
  \right)^{\alpha-1} \left( 2 - \sqrt{2} \: \xi \right),\\ \left[ 2 +
  \left( \sqrt{2} - \sqrt{1+\frac{1}{2\alpha}} \right) \xi \right]
  \xi^{2\alpha} &= \omega_- \left( -\frac{a}{R} \right)^{\alpha-1}
  \left( 2 + \sqrt{2} \: \xi \right),
\end{split}
\label{2.47}
\end{equation}
where
\begin{equation}
\omega_{\pm} = \frac{2\alpha}{1 \pm \left( 1 - \dfrac{1}{4\alpha^2}
\right)^{1/4}}.
\label{2.48}
\end{equation}
In the limit of $\alpha\to +\infty$ we have $\omega_+\to
\alpha$ and $\omega_-\to 2^5 \alpha^3$.

For the sake of clarity of the following analysis we assume $2\alpha$
to be an integer. Thus, for $R \gg a$ and large $\alpha$ one can find
approximate solutions to Eqs.~(\ref{2.47}):
\begin{equation}
\begin{split}
  &\xi_{p1} \approx \left(1 - \frac{1}{4\alpha}
    \sqrt{1+\frac{1}{2\alpha}} \: \xi_{p1}^{(0)} \right) \xi_{p1}^{(0)},\\
  &\phantom{+++++}\mathrm{with} \;\; \xi_{p1}^{(0)} = \omega_+^{1/2\alpha} \left(
    \sqrt{\frac{a}{R}} \right)^{1-1/\alpha} e^{i\pi (1+p/\alpha)},\\ &\xi_{p2} \approx
  \left(1 + \frac{1}{4\alpha}
    \sqrt{1+\frac{1}{2\alpha}} \: \xi_{p2}^{(0)} \right) \xi_{p2}^{(0)},\\
  &\phantom{+++++}\mathrm{with} \;\; \xi_{p2}^{(0)} = \omega_-^{1/2\alpha} \left(
    \sqrt{\frac{a}{R}} \right)^{1-1/\alpha} e^{i\pi (1+p/\alpha)},\\ &\xi_{3}^{(\pm)}
  \approx \pm 2 \left( \sqrt{2}-\sqrt{1+\dfrac{1}{2\alpha}} \right)^{-1},
\end{split}
\label{2.49}
\end{equation}
where $p=1,2,\ldots,2\alpha$. Each of these $4\alpha+2$ values of
$\xi$ defines a line of poles of $\Omega_\mathrm{S}(E)$ in the complex
$\kpp$-plane in accordance with
\begin{equation}
\begin{split}
  &\kpp_{n,p,j} = \frac{2\pi n + \pi + \arg\xi_{p,j}}{R} - \frac{i}{2 R} \ln
  \frac{R |\xi_{p,j}|^2}{a},\\ &\phantom{++++++++++++++++++}
  \mathrm{with} \;\; j=1,2;\\ &\kpp_{n,3}^{(\pm)} = \frac{2\pi n + \pi +
    \arg\xi_3^{(\pm)}}{R} - \frac{i}{2 R} \ln \frac{R |\xi_3^{(\pm)}|^2}{a}.
\end{split}
\label{2.50}
\end{equation}

\begin{figure}[h]
  \centerline{\epsfig{figure=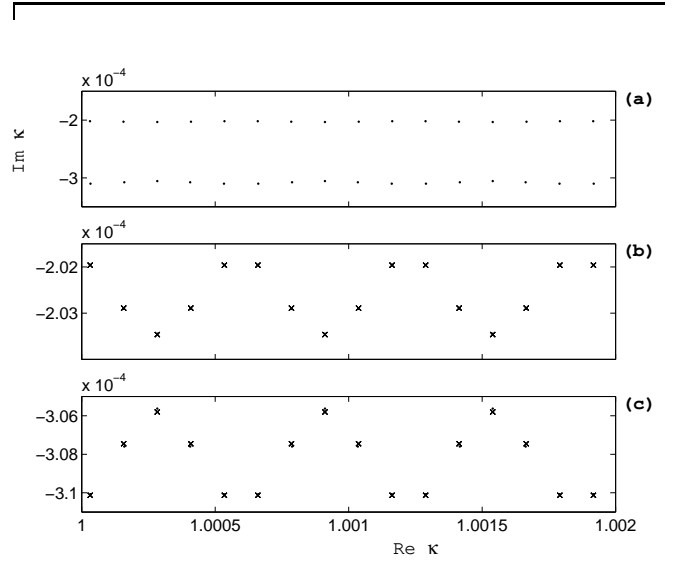,width=3.5in}}
  \caption{(a) First two ``bands'' of poles for the case of $\alpha=5/2$
    and $R/a = 10^4$; (b) magnification of the first band; (c)
    magnification of the second band. Dots correspond to (exact)
    values of the poles numerically computed directly from
    Eq.~(\ref{2.47}), while crosses show the same poles approximated
    by Eq.~(\ref{2.49}).}
\label{fig-5}
\end{figure}

The poles given by Eqs.~(\ref{2.50}) come in three ``bands''. The
first and the second bands of poles, corresponding to $\xi_{p,1}$ and
$\xi_{p,2}$ respectively, are essential for the wave packet dynamics,
while the third band, given by $\xi_3^{(\pm)}$, is completely screened
by the first two due to the relation $|\xi_1| < |\xi_2| \ll |\xi_3|$.
Figure~\ref{fig-5} shows the first two bands of poles for the
three-disk isosceles billiard with $\alpha = 5/2$ and $R/a = 10^4$.
Figure~\ref{fig-5}a displays the bands over some interval of real
$\kappa$-axis, while fig.~\ref{fig-5}b and fig.~\ref{fig-5}c magnify
the first and the second bands respectively. The dots in the last two
figures represent poles numerically computed directly from
Eqs.~(\ref{2.47}), and thus should be thought as of ``exact'' poles,
while crosses are the poles approximated by Eq.~(\ref{2.49}). In most
of the cases the dots and the crosses fall on top of each other, so
that one can conclude that Eqs.~(\ref{2.49}) accurately locate poles
of the autocorrelation function.

As we saw earlier, the size of the gap separating the poles and the
real $\kpp$-axis determines time decay of the envelope of the
autocorrelation function. Thus, the autocorrelation function envelope
decay, $C(t) \sim \exp(-\gamma_\alpha^{(3)} t)$, for the isosceles
three-disk billiard is governed by the rate
\begin{equation}
\gamma_\alpha^{(3)} \approx \dfrac{v}{R} \ln \left( \dfrac{R
}{a} \left|\xi_{p1}^{(0)}\right|^2 \right) = \dfrac{v}{\alpha R} \ln
\dfrac{2\alpha R}{\left[ 1 + \left( 1-\dfrac{1}{4\alpha^2}
\right)^{1/4} \right] a}.
\label{2.51}
\end{equation}
In the limit $\alpha \gg 1$ the decay rate becomes
$\gamma_\alpha^{(3)} \approx (v/\alpha R)\ln (\alpha R/a)$.

\begin{figure}[h]
  \centerline{\epsfig{figure=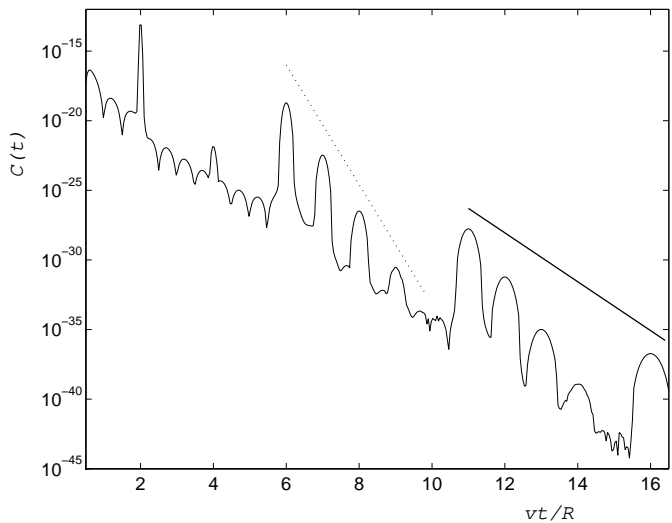,width=3.5in}}
  \caption{The autocorrelation function as a function of time for the
    isosceles three-disk billiard with $\alpha=5/2$. Disks ``1'' and ``2''
    are separated by distance $R=10^4a$. The wave packet of the de
    Broglie wavelength $\la=10^{-2}a$ is initially located as shown in
    fig.~\ref{fig-3} with $r_1=r_2=R/2$. The dotted and the solid
    straight lines represent $e^{-\lambda^{(2)} t}$ and $e^{-\gamma_\alpha^{(3)} t}$
    decays respectively.}
\label{fig-6}
\end{figure}

The time-domain autocorrelation function, $C(t)\approx
|\Omega_\mathrm{S}(t)|^2$, is now calculated using Eq.~(\ref{2.28}), with
$\Omega_\mathrm{S}(E)$ given by Eq~(\ref{2.46}). As before, the complex
$\kappa$-plane contour integration is performed by calculating the
residues of $\Omega_\mathrm{S}(E)$ and computing the sum over the poles
given by Eq.~(\ref{2.50}) with the index $n\in
(n_0-[R/2\sigma],n_0+[R/2\sigma])$. Here $2\pi n_0/R=k_0$ and the
square brackets denotes the integer part. The autocorrelation function
$C(t)$ for the three-disk isosceles billiard is shown in
fig.~\ref{fig-6} for the case of $\alpha=5/2$. The system is
characterized by $R/a=10^4$; the wave packet of the de Broglie
wavelength $\la=10^{-2}a$ is initially placed in the middle between
disks ``1'' and ``2'', see fig.~\ref{fig-3} with $r_1=r_2=R/2$.
Figure~\ref{fig-6} confirms our earlier predictions that the overall
envelope of the autocorrelation function decays as
$e^{-\gamma_\alpha^{(3)} t}$ with the decay exponent
$\gamma_\alpha^{(3)}$ given by Eq.~(\ref{2.51}). The trend of this
exponential decay is presented by the solid lines in the figure. The
dotted trend line corresponds to $e^{-\lambda^{(2)} t}$ decay, with
the rate $\lambda^{(2)}$, Eq.~(\ref{2.33}), being the two-disk
Lyapunov exponent of the unstable periodic orbit trapped between disks
``1'' and ``2''.

At first glance the autocorrelation function $C(t)$ appears merely as
some complicated sequence of decaying peaks. Nevertheless simple
physics underlies its structure. The peaks of the autocorrelation
function, or the wave packet partial revivals, occur at instants of
time at which the counterpart classical particle returns to its
initial point in phase space. Then, relatively larger peaks result
from the phase space trajectories with the smaller number of
collisions per unit time interval. It is because at every collision
event a dominant part of particle's probability density completely
escapes the billiard, and only a tiny part of this density proceeds to
the next collision in order to eventually contribute to the
autocorrelation function. In other words, phase space periodic orbits
with longer mean free paths result in stronger reconstruction peaks.
In the isosceles three-disk billiard with $\alpha > 1$ the long free
flight path trajectories are the ones that pass through disk ``3''
every second collision, see fig.~\ref{fig-3}, and return to the
initial point at times $t_n=(2\alpha n + 1)R/v$ with $n=1,2,\ldots$.
Thus, $C(t)$ exhibits relatively strong peaks at times $t_n$
corresponding to the scattering sequences ``132'', ``13232'',
``13132'' etc. During time intervals between the large peaks,
i.e. $t_{n-1} < t<t_n$, smaller wave packet reconstruction peaks occur
due to periodic orbits with shorter mean free paths, which are
determined by trajectories bouncing mostly between disks ``1'' and
``2''. This is how the two-disk Lyapunov exponent $\lambda^{(2)}$
enters the description of $C(t)$ for the three-disk billiard. Thus,
for times $t < t_1$ only the two-disk collision sequences ``12'',
``1212'' etc., contribute to the autocorrelation function, resulting
in the $e^{-\lambda^{(2)} t}$ decay.

Another distinctive feature of fig.~\ref{fig-6} is the absence of
revival peaks for times $R/v$, $3R/v$ and $5R/v$. That is because for
a classical particle moving with velocity $v=\hbar k_0/m$ there are no
phase space periodic orbits corresponding to these times. We have
already encountered same phenomenon while discussing the
autocorrelation function decay in the three-disk equilateral billiard.

Now it is also apparent how the two-disk decay $e^{-\lambda^{(2)} t}$
is recovered if disk ``3'' is removed to infinity, see
fig.~\ref{fig-3}. In the limit $\alpha \to \infty$ the time
$t_1$ of the first large reconstruction peak goes to infinity,
resulting in the two-disk billiard decay of the autocorrelation
function for $t < t_1 \to \infty$.

\bigskip

Before closing this section we give a brief summary of the results
obtained. We have used the multiple collision expansion technique to
provide a detailed first-principle calculation of the time dependent
autocorrelation function, $C(t)$, for quantum wave packets moving in
billiards composed of two and three hard disk scatterers. By
analytically constructing $C(t)$ for three-disk isosceles billiards we
have broadened the class of three-disk systems earlier treated by
similar methods \cite{gasp-II}. The applicability limits of these
methods were shown to be given by the condition of the high-energy
diffraction regime, Eq.~(\ref{2.27}). We found the decay of the
autocorrelation function for the case of a three-disk equilateral
billiard to be mainly exponential apart from a regular sequence of
wave packet reconstruction peaks of equal relative strength. On the
other hand, $C(t)$ decays non-uniformly in the case of a three-disk
isosceles system, and features different decay rates at different time
scales. Thus, at times shorter that $(2\alpha+1)R/v$, see
fig.~\ref{fig-3}, the decay is entirely determined by the Lyapunov
exponent of the shortest two-disk classical periodic orbit, while at
long times, the overall envelope decays at a slower rate given by
Eq.~(\ref{2.51}). A well pronounced structure of revival peaks is
again observed in the autocorrelation function of the three-disk
isosceles billiard. As we will show in the Section III the relative
strengths of the peaks are determined by the number of interfering
periodic orbits of the counterpart classical system.

\section{Semiclassical description}

In this section we present a simple method for predicting relative
strength of the peaks of the wave packet autocorrelation function
$C(t)$, see Eq.~(\ref{intr-1}), based on the semiclassical Van Vleck
propagator \cite{gutz}. The semiclassical analysis of the
autocorrelation function was earlier performed by Tomsovic and Heller
for the case of the stadium billiard \cite{toms-2}. We start with
applying the semiclassical method for two- and three-disk billiards
studied in the previous section, and then, use it to investigate such
more complicated scattering systems as the three-disk generic billiard
and the two-, three- and four-sphere billiards in three spatial
dimensions.

\subsubsection{Semiclassical approach}

In the limit of short de Boglie wavelengths the time evolution of a
quantum state in a hard-disk or hard-sphere billiard can be described
by the semiclassical Van Vleck propagator~\cite{gutz}
\begin{equation}
\begin{split}
  &G_\mathrm{sc}(\vr,\vr';t) \equiv \langle \vr | G_\mathrm{sc}(t) | \vr' \rangle\\ &=
  \left( \frac{1}{2\pi i\hbar} \right)^{d/2} \sum_{\tilde{\gamma}}
  D_{\tilde{\gamma}}^{1/2} \exp \left( i\frac{S_{\tilde{\gamma}}(\vr,\vr';t)}{\hbar}
    - i\frac{\pi \nu_{\tilde{\gamma}}}{2} \right).
\end{split}
\label{3.1}
\end{equation}
The summation in this expression goes over all classical paths
$\tilde{\gamma}$ connecting points $\vr'$ and $\vr$ (in
$d$-dimensional space) in time $t$. $S_{\tilde{\gamma}}(\vr,\vr';t)$
represents the classical action along the path $\tilde{\gamma}$, and
$\nu_{\tilde{\gamma}}$ is an index equal to twice the number of
collisions of the particle with hard scatterers during time $t$
\cite{gasp-II}. The Van Vleck determinant, $D_{\tilde{\gamma}} =
|\det(-\partial^2 S_{\tilde{\gamma}}/\partial\vr\partial\vr')|$,
corresponds, up to an appropriate normalization factor, to the
classical probability of the path $\tilde{\gamma}$ \cite{peres-3}.

The autocorrelation overlap due to the propagator given by
Eq.~(\ref{3.1}) and the an initial quantum state $|\phi_0\rangle$ can
be written as
\begin{equation}
\langle \phi_0 | G_\mathrm{sc}(t) | \phi_0 \rangle = \int d\vr \int
 d\vr' \phi_0^*(\vr) G_\mathrm{sc}(\vr,\vr';t) \phi_0(\vr'),
\label{3.2}
\end{equation}
where asterix denotes complex conjugate. Let us now assume that the
wave function $\phi_0(\vr)$ is localized about point $\vR_0$ and has
an average momentum $\hbar \vk_0$,
\begin{equation}
\phi_0(\vr) = |\phi_0(\vr)| e^{i\vk_0 \vr}.
\label{3.3}
\end{equation}
Since both wave functions in the integrand on the right hand side of
Eq. (\ref{3.2}) are localized about $\vR_0$, we expand the action
$S_{\tilde{\gamma}}(\vr,\vr';t)$ about the points $\vr = \vR_0$ and $\vr'
=\vR_0$ \footnote{The small parameter in the above expansion is the
  ratio of spatial distances of $\vr,\vr'$ from $\vR_0$ to the total
  distance travelled by a classical particle in the periodic orbit.},
as
\begin{equation}
S_{\tilde{\gamma}}(\vr,\vr';t) \approx S_\gamma(\vR_0,\vR_0;t) +
\hbar\vk_\gamma (\vr-\vR_0) - \hbar\vk'_\gamma (\vr'-\vR_0),
\label{3.4}
\end{equation}
where a classical particle traveling along the closed path $\gamma$
would leave the point $\vR_0$ with momentum $\hbar\vk'_\gamma =
-\partial S_\gamma(\vR_0,\vr';t)/\partial\vr' |_{\vr'=\vR_0}$ and
return to $\vR_0$ after time $t$ having momentum $\hbar\vk_\gamma =
\partial S_\gamma(\vr,\vR_0;t)/\partial\vr
|_{\vr=\vR_0}$.
Substitution of Eq.~(\ref{3.1}) along with
Eqs.~(\ref{3.3}) and (\ref{3.4}) into Eq.~(\ref{3.2}) yields
\begin{equation}
\begin{split}
  &\langle \phi_0 | G_\mathrm{sc}(t) | \phi_0 \rangle \approx \left( \frac{1}{2\pi i\hbar}
  \right)^{d/2} \sum_\gamma D_\gamma^{1/2}\\ &\times \exp \left(
    i\frac{S_\gamma(\vR_0,\vR_0;t)}{\hbar} - i\frac{\pi \nu_\gamma}{2} \right)
  [\bar{\phi}_0(\vk_\gamma)]^* \bar{\phi}_0(\vk'_\gamma),
\end{split}
\label{3.5}
\end{equation}
where
\begin{equation}
\bar{\phi}_0(\vk) = e^{i\vk \vR_0} \int d\vr |\phi_0(\vr)|
e^{i(\vk_0-\vk)\vr}.
\label{3.6}
\end{equation}
In deriving Eq.~(\ref{3.5}) we used $D_{\tilde{\gamma}} \approx
D_\gamma$ and $\nu_{\tilde{\gamma}}=\nu_\gamma$ assuming that the
paths, $\tilde{\gamma}$ and $\gamma$, are close and follow the same
collision sequence.

Let us now assume the function $|\phi_0(\vr)|$ to be sufficiently smooth
in order for $\bar{\phi}_0(\vk)$, defined by Eq.~(\ref{3.6}), to be
sharply peaked about $\vk_0$. Then, the main contribution to the
autocorrelation amplitude in Eq.~(\ref{3.5}) comes from paths $\gamma$ with
$\vk_\gamma \approx \vk'_\gamma \approx \vk_0$.  Thus, in order to obtain a leading
contribution to $\langle \phi_0 | G_\mathrm{sc}(t) | \phi_0 \rangle$ at a {\it fixed}
time $t$, one can restrict the summation in Eq.~(\ref{3.5}) only to
classical periodic orbits $\bar{\gamma}$ passing through a small
neighborhood of the phase space point $(\vR_0,\vk_0)$, and therefore
having the momentum $p \approx \hbar k_0$. The classical actions along such
periodic orbits can be written as $S_{\bar{\gamma}} \approx (\hbar^2 k_0^2/2m) \, t$,
and are the same for all $\bar{\gamma}$'s.  The autocorrelation amplitude
then reads
\begin{equation}
\begin{split}
  \langle \phi_0 | G_\mathrm{sc}(t) | \phi_0 \rangle &\approx e^{i\frac{\hbar t}{2m} k_0^2} \left(
    \frac{1}{2\pi i\hbar} \right)^{d/2}\\ &\times \left( \int d\vr |\phi_0(\vr)|
  \right)^2 \sum_{\bar{\gamma}} D_{\bar{\gamma}}^{1/2}
  e^{-i\frac{\pi\nu_{\bar{\gamma}}}{2}}.
\end{split}
\label{3.7}
\end{equation}
Equation~(\ref{3.7}) is suitable only for predicting the values of the
autocorrelation function $C(t)\approx |\langle |\phi_0 | G_\mathrm{sc}(t) | \phi_0 \rangle|^2$
at times $t=t_{\bar{\gamma}}$ such that there exists at least one
classical periodic trajectory $\bar{\gamma}$ of period $t_{\bar{\gamma}}$
passing through the spatial point $R_0$ with momentum $\hbar k_0$. Due to
the narrow momentum distribution of the initial wave packet, given by
the function $\bar{\phi}_0(\vk)$ in Eq.~(\ref{3.5}), $C(t)$ decreases by
orders of magnitude as the time $t$ changes to a value such that all
the classical periodic orbits of period $t$ have momenta different
from $\hbar k_0$. Therefore, we come to a conclusion that the time decay
of the autocorrelation function consists of sequence of sharp peaks
centered at the times $t_{\bar{\gamma}}$. Equation~(\ref{3.7}) is then
only suitable for predicting the relative strength and time-location
of these peaks.

It will be shown below that the probability measure of periodic orbits
(or $D_{\bar{\gamma}}$) decreases exponentially with the increase of the
number of collisions a classical particle undergoes while traveling
along these orbits.  Therefore, the value of the autocorrelation
function at a given peak at time $t$ is predominantly determined by a
subset $\{\gamma'\}$ of the set of all classical periodic trajectories
$\{\bar{\gamma}\}$ of length $vt=(\hbar k_0/m)t$, such that the members of the
subset have the {\it smallest} number of scattering events, $N(t)$,
possible for time $t$.  Indexes $\nu_{\gamma'}=2N(t)$ are the same for all
members of the subset $\{\gamma'\}$, yielding
\begin{equation}
C(t) \sim \left( \sum_{\gamma'} \sqrt{D_{\gamma'}} \right)^2.
\label{3.8}
\end{equation}
Equation~(\ref{3.8}) allows one to predict the relative magnitude of
peaks of the autocorrelation function $C(t)$ by searching
(analytically or numerically) for the periodic orbits with the
smallest number of scattering events during a given time $t$. We will
employ this formula in the sequel to calculate the autocorrelation
function decay for wave packets in various hard-disk and hard-sphere
billiards.

If properly modified, the above technique is also applicable for
calculation of the peaks of the {\it classical} autocorrelation
function $C_\mathrm{cl}(t)$, {i.e.} a fraction of classical
trajectories in a small phase space region around the initial location
of a classical particle, which return to this region after time
$t$. The classical autocorrelation function gives the phase space
return probability for a classical particle described by a phase space
distribution function rather by the exact coordinates. It
characterizes the escape of classical trajectories from a small
neighborhood of a chaotic repeller of the system. In mixing chaotic
systems, such as hard-disk and hard-sphere billiards, the
autocorrelation function decays exponentially with time,
$C_\mathrm{cl}(t) \sim e^{-\gamma_\mathrm{cl} t}$, with the decay rate
$\gamma_\mathrm{cl}$ known as the escape rate on the repeller
\cite{gasp-II,kantz}. The changes one needs to introduce in
Eq.~(\ref{3.8}) are apparent. Due to the absence of interference in
classical mechanics one has to directly sum the probabilities of the
periodic orbits, $D_{\gamma'}$, rather than the probability amplitudes
$\sqrt{D_{\gamma'}}$. This leads to
\begin{equation}
C_\mathrm{cl}(t) \sim \sum_{\gamma'} D_{\gamma'}.
\label{3.9}
\end{equation}

Finally, one needs to specify quantities $D_{\gamma'}$ entering the
RHS of Eqs.~(\ref{3.8}) and (\ref{3.9}). This can be done taking into
account that the Van Vleck determinant $D_{\gamma'}$ equals, up to a
normalization factor, the classical probability measure of the
trajectory $\gamma'$, e.g. see \cite{peres-3}. Thus, in
$d$-dimensional space the probability density to find a particle at a
distance $R$ from the source radiating particles with some initial
velocity distribution is proportional to $1/R^{d-1}$. Furthermore, the
probability for a particle to undergo a collision with a scatterer in
a given direction is described by the differential cross section
$\sdif(\theta)$, with $\theta$ being the scattering angle. Therefore,
the probability for the particle to follow a periodic collision
sequence $\eta_{\gamma'}=\{i,j,\ldots,q,r,s\}$, constituting an orbit
$\gamma'$, is
\begin{equation}
\begin{split}
  D_{\gamma'} \sim \frac{\sdif(\pi-\phi_i) }{R_{ij}^{d-1}}& \ldots \frac{\sdif(\pi-\phi_q)
  }{R_{qr}^{d-1}}\\ &\times \frac{\sdif(\pi-\phi_r) }{R_{rs}^{d-1}}
  \frac{\sdif(\pi-\phi_s) }{R_{si}^{d-1}},
\end{split}
\label{3.10}
\end{equation}
where $R_{ij}$ is the center-to-center separation between scatterers
$i$ and $j$, and $\phi_i$ is the angle of the periodic orbit polygon
(with the vertices at the scatterer centers) corresponding to the
$i^\mathrm{\, th}$ vertex. The separation distances satisfy an obvious
relation,
\begin{equation}
R_{ij} + \ldots + R_{qr} + R_{rs} + R_{si} = vt,
\label{3.11}
\end{equation}
which implicitly provides the time dependents to the RHS of
Eqs.~(\ref{3.8}) and (\ref{3.9}).

As we will see below, the expression for the quantum (classical)
autocorrelation function proposed in this section, despite its
simplicity, accurately predicts times and relative magnitudes of the
wave packet (distribution function) reconstruction peaks.
Nevertheless, application of more sophisticated techniques, like the
one presented in the previous chapter, is required if one needs to
obtain absolute (and not relative) values of the autocorrelation
function for a wide range of times, including time intervals between
the neighboring peaks. One needs to have detailed knowledge of the
particle's wave function in order to predict $C(t)$ for times $t$
other than the peak times, i.e. for times that have no phase space
period orbits of velocity $v$ corresponding to them. This requires the
construction of the full quantum propagator for a given system by
methods analogous to the one presented in Section II.

\subsection{Application to studied cases}

We first start with applying the semiclassical technique to calculate
the peaks of the autocorrelation function for the hard-disk scattering
systems analyzed in Section II by the method of multiple collision
expansions, i.e. for the two-disk billiard, and three-disk equilateral
and isosceles billiards. After that we will treat such more
complicated system as the tree-disk generic billiard, as well as some
hard-sphere billiards in three dimension.

\subsubsection{Two-disk billiard}

In the case of the two-disk scattering system, see fig~\ref{fig-1},
there is only one scattering sequence, ``1212$\ldots$'', contributing
to a peak of the autocorrelation function, $C(t_n)$, at time
$t_n=2nR/v$, with $n=1,2,\ldots$. The probability weight $D_n$,
corresponding to a periodic orbit of length $vt_n$, is calculated
according to Eq.~(\ref{3.10}) with $d=2$ and the semiclassical
hard-disk differential cross section
\begin{equation}
\sdif(\theta) = \frac{a}{2} \left| \sin\frac{\theta}{2} \right|,
\label{3.12}
\end{equation}
where $\theta=\pi$ for the back scattering. Then,
\begin{equation}
C(t_n) \sim D_n \sim \left( \frac{a}{2R} \right)^{2n} =
\exp\left(-\lambda^{(2)} t_n \right),
\label{3.13}
\end{equation}
where $\lambda^{(2)}$ is the two-disk Lyapunov exponent given by
Eq.~(\ref{2.33}).

Since the right hand sides of Eqs.~(\ref{3.8}) and (\ref{3.9}) reduce
to the return probability $D_n$ of a single collision sequence, we see
that the classical autocorrelation function decays exactly in the same
manner as the quantum one:
\begin{equation}
C_\mathrm{cl}(t_n) \sim \exp\left(-\lambda^{(2)} t_n \right).
\label{3.14}
\end{equation}
As we will see later, this similarity is the consequence of the fact
that the Kolmogorov-Sinai entropy for the two-disk periodic orbit is
zero \cite{gasp-I}, {i.e.} there is no information production in the
system since for any time $t_n=2nR/v$ there exists only one trajectory
leading to the wave packet (distribution function) partial
reconstruction. Thus, the phenomenon of interference between different
trajectories is absent in the quantum case, resulting in the same
escape rates for classical and semiclassical particles.

\subsubsection{Three-disk equilateral billiard}

\begin{figure}[h]
  \centerline{\epsfig{figure=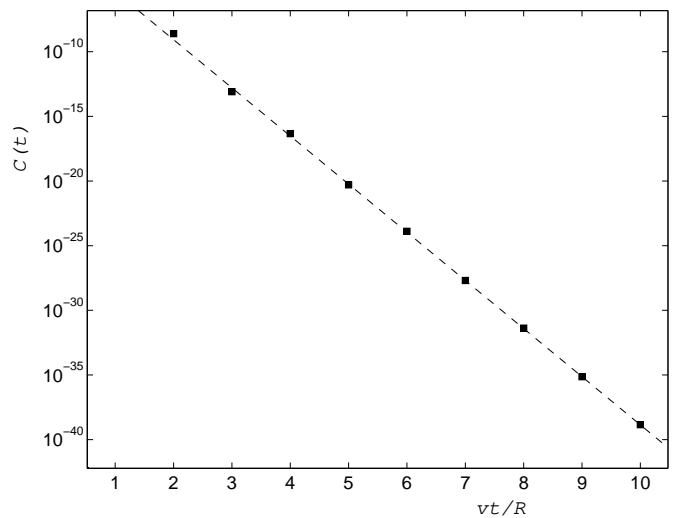,width=3.5in}}
  \caption{Peaks of the autocorrelation function for the equilateral
    three-disk billiard calculated in accordance with
    Eqs.~(\ref{3.8}), (\ref{3.10}) and (\ref{3.12}). The dashed line
    shows $e^{-\gamma^{(3)} t}$ decay, with $\gamma^{(3)}$ given by
    Eq.~(\ref{2.43}). The radii of the disks constituting the billiard
    equal $a=1$, and the disk center-to-center separation is $R=10^4$.
    This figure is to be compared with fig.~\ref{fig-4}.}
\label{fig-7}
\end{figure}

Let us now address the three-disk scattering system, with the
scatterers centered in the vertices of an equilateral triangle.
Figure~\ref{fig-7} shows the autocorrelation peaks as a function of
time, which were computed numerically according to Eqs.~(\ref{3.8}),
(\ref{3.10}) and (\ref{3.12}) by summing over all collision sequences
satisfying Eq.~(\ref{3.11}). The dashed trend line in the figure
represents the exponential decay with the rate $\gamma^{(3)}$ given by
Eq.~(\ref{2.43}). The billiard is characterized by the disk radius
$a=1$, and the disk center-to-center separation $R=10^4$. The system
is identical to the billiard considered in Section II, see
fig.~\ref{fig-1}, and the decay of the autocorrelation peaks is to be
compared with the one presented in fig~\ref{fig-2}. Note, that the de
Broglie wavelength $\la$ does not enter the sum in Eq.~(\ref{3.8}),
being contained in a possible prefactor, and is therefore not
important for determining the {\it relative} strength of the peaks.
Figure~\ref{fig-7} clearly shows that, as in the two-disk case, the
semiclassical theory recovers the autocorrelation function decay rate
obtained in Section II by means of the multiple collision expansion
technique.

The decay rate $\gamma^{(3)}$ given by Eq.~(\ref{2.43}) for the
equilateral three-disk billiard can be exactly recovered using the
semiclassical method. In order to calculate $\gamma^{(3)}$ one needs
to sum probability amplitudes $\sqrt{D_{\gamma'}}$ over all collision
sequences $\eta_{\gamma'}$ satisfying Eq.~(\ref{3.11}). Once again,
this can be accomplished with the help of the matrix method
\cite{gasp-II} discussed in the previous section. We construct a
one-collision transition matrix $\vq$ according to
\begin{equation}
\vq =
\begin{array}{cc}
  ^{1\cdot 2} \; ^{1\cdot 3} \; ^{2\cdot 1} \; ^{2\cdot 3} \;
  ^{3\cdot 1} \; ^{3\cdot 2} & \\ \left(
\begin{array}{cccccc}
0 & 0 & x & w & 0 & 0\\ 0 & 0 & 0 & 0 & x & w\\ x & w & 0 & 0 & 0 &
0\\ 0 & 0 & 0 & 0 & w & x\\ w & x & 0 & 0 & 0 & 0\\ 0 & 0 & w & x & 0
& 0
\end{array}\right) &
\begin{array}{c}
\!\! ^{1\cdot 2}\\
\!\! ^{1\cdot 3}\\
\!\! ^{2\cdot 1}\\
\!\! ^{2\cdot 3}\\
\!\! ^{3\cdot 1}\\
\!\! ^{3\cdot 2}
\end{array}
\end{array}
\label{3.15}
\end{equation}
with
\begin{equation}
x \equiv \left( \frac{a}{2R} \right)^{1/2} \;\;\;\;\;\;\;\;
\mathrm{and} \;\;\;\;\;\;\;\; w \equiv \left( \frac{\sqrt{3}a}{4R}
\right)^{1/2}.
\label{3.16}
\end{equation}
Here, $x$ and $w$ are the values of the amplitude
$\sqrt{\sdif(\pi-\phi)/R}$, with $\sdif$ given by Eq.~(\ref{3.12}) and
$\phi$ taking values of $0$ and $\pi/3$ respectively. The matrix $\vq$
describes a transition due to a single collision event in the
six-dimensional space spanned by directions $(1\!\to\!  2)$,
$(1\!\to\!  3)$, $(2\!\to\! 1)$, $(2\!\to\!
3)$, $(3\!\to\! 1)$ and $(3\!\to\! 2)$. This matrix
allows one to express the sum of overlap amplitudes in
Eq.~(\ref{3.18}) for times $t_n=n R/v$, with number of collisions
$n=2,3,\ldots$, according to
\begin{equation}
\sum_{\gamma'} D_{\gamma'}^{1/2}(t_n) = \left( \vq^n \right)_{1,1},
\label{3.17}
\end{equation}
where the subscript in the RHS denotes that the one-one element of the
matrix is taken. The autocorrelation function at the $n^{\,
\mathrm{th}}$ collision is related to the one at the $(n
+1)^{\,\mathrm{th}}$ collision by
\begin{equation}
\frac{C(t_n+R/v)}{C(t_n)} = \left( \frac{\left( \vq^{n +1}
  \right)_{1,1}}{\left( \vq^n \right)_{1,1}} \right)^2.
\label{3.18}
\end{equation}
For large number of collisions, $n \gg 1$, the largest eigenvalue of
the matrix $\vq$, equal to $x+w$, dominates both the numerator and the
denominator of Eq.~(\ref{3.18}), so that
\begin{equation}
\begin{split}
  &\lim_{t_n\to +\infty} \frac{C(t_n+R/v)}{C(t_n)} = (x+w)^2\\ &= \frac{a}{R}
  \left[ \left(1/2\right)^{1/2}+\left(3^{1/2}/4\right)^{1/2} \right]^2
  = \exp \left( -\gamma^{(3)} \frac{R}{v} \right),
\end{split}
\label{3.19}
\end{equation}
where $\gamma^{(3)}$ is the decay rate given by Eq.~(\ref{2.43}).

Finally let us mention that the classical escape rate
$\gamma_\mathrm{cl}^{(3)}$ can be obtained with the help of the transition
matrix $\vq$, given by Eq.~(\ref{3.15}), if one redefines $x$ and $w$
by replacing the square roots in Eq.~(\ref{3.16}) by the first powers,
i.e.
\begin{equation}
x\to \frac{a}{2R} \;\;\;\;\; \mathrm{and} \;\;\;\;\;
w\to \frac{\sqrt{3}a}{4R}.
\label{3.20}
\end{equation}
Then,
\begin{equation}
\begin{split}
  \frac{C_\mathrm{cl}(t_n+R/v)}{C_\mathrm{cl}(t_n)} &= \frac{\left(
      \vq^{n+1} \right)_{1,1}}{\left( \vq^n \right)_{1,1}} \;
  \stackrel{t\to \infty}{\longrightarrow} x+w\\ &= \frac{a}{R} \left[ \frac{1}{2} +
    \frac{\sqrt{3}}{4} \right] = \exp \left( -\gamma_\mathrm{cl}^{(3)}
    \frac{R}{v} \right),
\end{split}
\label{3.21}
\end{equation}
with the classical escape rate given by
\begin{equation}
\gamma_\mathrm{cl}^{(3)} = \frac{v}{R} \ln \frac{4R}{\left[ 2 +
\sqrt{3} \right] a} \approx \frac{v}{R} \ln \frac{1.07 \, R}{a}.
\label{3.22}
\end{equation}
One can see that the absence of interference in the classical case
results in faster particle escape from the scattering system. The
classical escape rate for the three-disk equilateral billiard was
first obtained by Gaspard and Rice \cite{gasp-I}.

\subsubsection{Three-disk isosceles billiard}

\begin{figure}[h]
  \centerline{\epsfig{figure=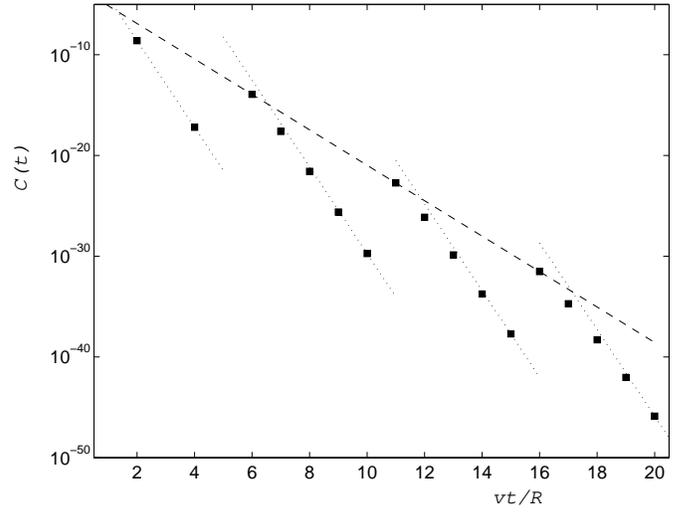,width=3.5in}}
  \caption{Peaks of the autocorrelation function as predicted by
    Eq.~(\ref{3.8}) for the isosceles three-disk billiard with
    $\alpha=5/2$. The dashed line shows $e^{-\gamma_\alpha^{(3)} t}$ decay, with
    $\gamma_\alpha^{(3)}$ given by Eq.~(\ref{2.51}), while the dotted lines show
    the trend of the $e^{-\lambda^{(2)} t}$ decay, with $\lambda^{(2)}$ defined by
    Eq.~(\ref{2.33}). The billiard is parametrized by $a=1$ and
    $R=10^4$. This figure is to be compared with fig.~\ref{fig-6}.}
\label{fig-8}
\end{figure}

In order to complete the comparison of predictions of the
semiclassical methods with the results of the detailed binary
collision expansion studies, we consider the case of the three-disk
billiard with the scatterers centered in the vertices of an isosceles
triangle. Figure~\ref{fig-8} displays the peaks of the autocorrelation
function in the system shown in fig.~\ref{fig-3} with $\alpha=5/2$,
$a=1$ and $R=10^4$. The structure of the decay is twofold. There are
relatively big recurrences of the wave packet at times $t_n= (2\alpha
n + 1)R/v = (5n+1)R/v$ with $n=1,2,3,\ldots$. The magnitudes of these
recurrences follow $e^{-\gamma_\alpha^{(3)} t}$ decay (represented by
the dashed line), with the decay rate $\gamma_\alpha^{(3)}$ given by
Eq.~(\ref{2.51}). In between any two large neighboring peaks the
autocorrelation function decays rapidly, approximately following
$e^{-\lambda^{(2)} t}$ decay (dotted lines), with the two-disk
Lyapunov exponent $\lambda^{(2)}$ defined by Eq.~(\ref{2.33}). The
sequence of autocorrelation function peaks in fig.~\ref{fig-8} is
almost identical to the one in fig.~\ref{fig-6}.

We will now use the matrix method to derive the autocorrelation
function envelope decay rate $\gamma_\alpha^{(3)}$ directly from
Eq.~(\ref{3.8}). As it was mentioned above the sum in the RHS of
Eq.~(\ref{3.8}) goes only over such periodic orbits $\gamma'$ that
comprise the smallest number of scattering events possible for a
periodic trajectory of length $vt$. This is because the overlap
amplitudes $\sqrt{D_{\gamma'}}$ corresponding to trajectories with the
longest mean free path are given by products of the smallest number of
$\sqrt{\sdif/R} \ll 1$ terms, and therefore are the dominant ones. In
the case of the three-disk isosceles billiard with $\alpha > 1$ these
long mean free path trajectories are the ones that pass through the
scatter ``3'' every second collision, see fig.~\ref{fig-3}. This
amounts to the collision sequence ``132'' at time $t_1 = (2\alpha +
1)R/v$, collision sequences ``13132'' and ``13232'' at time $t_2 =
(4\alpha + 1)R/v$, etc. In general there are $2^{n-1}$ different
periodic orbits of length $vt_n = (2n\alpha + 1)R$ contributing to the
autocorrelation function peak $C(t_n)$.

The relative probability weights of the above trajectories are
determined according to Eq.~(\ref{3.10}) with the differential cross
section given by Eq.~(\ref{3.12}). Thus, collisions $(1\!\to
\!  3\!\to \! 1)$ and $(2\!\to \!  3\!\to \!
2)$ are described by the cross section $\sdif_3 = a/2$, while
$(1\!\to \! 3\!\to \! 2)$ and $(2\!\to \!
3\!\to \! 1)$ by $\sdif'_3 = (a/2) \cos(\phi_3/2) = (a/2)
\sqrt{1-1/4\alpha^2}$, where $\phi_3$ denotes the angle of the
triangle corresponding to vertex ``3'', see fig.~\ref{fig-3}. The
first and the last collisions of every periodic trajectory deflects
the moving particle by the angle $\theta = \pi - \phi_1 = \pi/2 +
\phi_3/2$, and is described by the cross section $\sdif_1 = \sdif_2 =
(a/2^{3/2}) \sqrt{1+1/2\alpha}$. Here $\phi_1$ and
$\phi_2$($=\!\phi_1$) are the other two angles of the isosceles
triangle satisfying $2\phi_1 + \phi_3 = \pi$. Thus, the sum of overlap
amplitudes at time $t_n = (2n\alpha + 1)R/v$, corresponding to
periodic orbits of $2n+1$ collisions, can be written as
\begin{equation}
\sum_{\gamma'} D_{\gamma'}^{1/2}(t_n) = \sqrt{\frac{\sdif_1}{\alpha
R}} \: \left( \vq^{2n-1} \right)_{1,2} \, \sqrt{\frac{\sdif_2}{R}},
\label{3.23}
\end{equation}
where
\begin{equation}
\vq =
\begin{array}{cc}
  ^{2\cdot 3} \; ^{3\cdot 1} \; ^{1\cdot 3} \; ^{3\cdot 2} & \\
  \left(
\begin{array}{cccc}
0 & w & 0 & x\\ 0 & 0 & x & 0\\ 0 & x & 0 & w\\ x & 0 & 0 & 0
\end{array}\right) &
\begin{array}{c}
\!\! ^{2\cdot 3}\\
\!\! ^{3\cdot 1}\\
\!\! ^{1\cdot 3}\\
\!\! ^{3\cdot 2}
\end{array}
\end{array}
\label{3.24}
\end{equation}
with
\begin{equation}
  x \equiv \sqrt{\frac{a}{2\alpha R}} \;\;\;\;\; \mathrm{and}
  \;\;\;\;\; w \equiv \sqrt{\frac{a}{2\alpha R}} \left(
    1-\frac{1}{4\alpha^2} \right)^{1/4}.
\label{3.25}
\end{equation}
Repeating the arguments used in the case of the three-disk equilateral
billiard we find that
\begin{equation}
\begin{split}
  \frac{C(t_{n+1})}{C(t_n)} &= \left( \frac{\left( \vq^{2n+1}
      \right)_{1,2}}{\left( \vq^{2n-1} \right)_{1,2}} \right)^2
  \stackrel{n\to \infty}{\longrightarrow} \: x^2 (x+w)^2\\ &= \left( \frac{a}{2\alpha R}
  \right)^2 \left[ 1 + \left( 1-\frac{1}{4\alpha^2} \right)^{1/4}
  \right]^2,
\end{split}
\label{3.26}
\end{equation}
since $\sqrt{x(x+w)}$ is the largest eigenvalue of the matrix
$\vq$. The time interval between any two successive large peaks of the
autocorrelation function is $t_{n+1}-t_n=2\alpha R/v$, so that for
$n\gg 1$ we get
\begin{equation}
C(t_{n+1}) \approx C(t_n) \exp \left[ - \gamma_\alpha^{(3)}
(t_{n+1}-t_n)\right],
\label{3.27}
\end{equation}
with the decay rate $\gamma_\alpha^{(3)}$ given by
Eq.~(\ref{2.51}). Once again we observe strong agreement between the
prediction of the semiclassical analysis and the results of the
diffraction regime approximation obtained in Section~II.

The classical decay rate $\gamma_{\mathrm{cl},\,\alpha}^{(3)}$ for the
three-disk isosceles billiard can be obtained by the following
modification of the elements of the transition matrix $\vq$:
\begin{equation}
  x \to \frac{a}{2\alpha R} \;\;\;\;\; \mathrm{and}
  \;\;\;\;\; w \to \frac{a}{2\alpha R}
  \sqrt{1-\frac{1}{4\alpha^2}} \: .
\label{3.28}
\end{equation}
Then,
\begin{equation}
\begin{split}
  \frac{C_\mathrm{cl}(t_{n+1})}{C_\mathrm{cl}(t_n)} &= \frac{\left(
      \vq_\alpha^{2n+1} \right)_{1,2}}{\left( \vq_\alpha^{2n-1} \right)_{1,2}}
  \stackrel{n\to \infty}{\longrightarrow} \: x (x+w)\\ &= \exp \left[ -
    \gamma_{\mathrm{cl},\,\alpha}^{(3)} (t_{n+1}-t_n)\right],
\end{split}
\label{3.29}
\end{equation}
with the classical decay rate
\begin{equation}
\gamma_{\mathrm{cl},\,\alpha}^{(3)} = \frac{v}{\alpha R} \ln
  \frac{2\alpha R}{\left[ 1 + \sqrt{1-\dfrac{1}{4\alpha^2}}
  \right]^{1/2} a}.
\label{3.30}
\end{equation}
Comparison of Eqs.~(\ref{2.51}) and (\ref{3.30}) shows that, as in the
case of the three-disk equilateral billiard, classical particle escape
in the isosceles billiard takes place at higher rate than the
corresponding quantum process.

\subsection{More billiards}

As we have seen, the semiclassical method presented here is suitable
for predicting the main features of the autocorrelation decay in
hard-disk scattering systems. We now apply the method to problems
which could not be easily treated by the technique of explicit
calculation of scattering resonances used in the Section II.

\subsubsection{Generic three-disk billiard}

The first system we address is a three-disk billiard of the most
general type: the disks of radii $a$ are centered in the vertices of a
triangle of unequal sides $R$, $\alpha R$ and $\beta R$, where for
concreteness we take $\beta > 1,\alpha$. In order to visualize the
system one should consider the three-disk isosceles billiard shown in
fig.~\ref{fig-3}, and elongate the side ``13'' of the triangle from
its original length $\alpha R$ to the new length $\beta R$. It is a
formidable problem to calculate the scattering resonances (and the
corresponding residues) governing the time evolution of a wave packet
for such a system. This makes the multiple collision expansion
technique, used to study the three-disk equilateral and isosceles
billiards, inefficient in the case of the generic three-disk
scattering system. On the other hand the semiclassical approach of
this section is quite easy to implement. It allows one to extract such
important information about the autocorrelation function as the
relative strength of the revival peaks and the overall envelope decay
rate.

The trajectories with the longest free flight path are the ones that
bounce most of their time between disks ``1'' and ``3'' with the
largest center-to-center separation $\beta R$. In the limit of a large
number of collisions, $n \gg 1$, there is a single periodic collision
sequence ``1313\ldots132'' resulting in a relatively strong wave
packet recurrence at times $t_n=[1+\alpha+(n-2)\beta] R/v$. In
accordance with Eqs.~(\ref{3.8}), (\ref{3.10}) and (\ref{3.12}) we
have
\begin{equation}
\begin{split}
  C(t_n) &\sim \left( \frac{a}{2\beta R} \right)^{n-2} \frac{a }{2R}
  \cos\left(\frac{\phi_2}{2}\right) \; \frac{a}{2\alpha R}
  \cos\left(\frac{\phi_3}{2}\right)\\ &\sim \exp \left( -\lambda_\beta^{(2)} t_n
  \right),
\end{split}
\label{3.31}
\end{equation}
where $\phi_2$ and $\phi_3$ are the triangle angles at vertices ``2''
and ``3'' respectively, and
\begin{equation}
\lambda_\beta^{(2)} = \frac{v}{\beta R} \ln \frac{2\beta R}{a}
\label{3.32}
\end{equation}
is the two-disk Lyapunov exponent, see Eq.~(\ref{2.33}), corresponding
to disks ``1'' and ``3''.

\begin{figure}[h]
  \centerline{\epsfig{figure=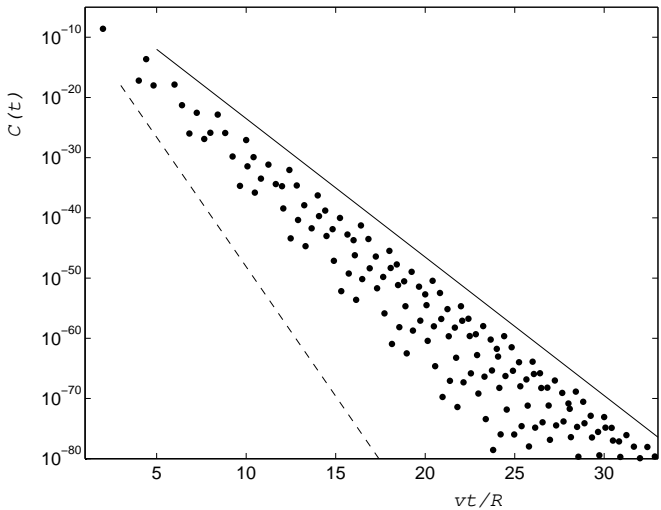,width=3.5in}}
  \caption{Peaks of the autocorrelation function for the three-disk
    billiard with $\alpha = \sqrt{2}$, $\beta = 2$, $a=1$ and $R=10^4$. Solid
    line represents $e^{-\lambda_\beta^{(2)} t}$ decay, with $\lambda_\beta^{(2)}$
    calculated according to Eq.~(\ref{3.32}); dashed line corresponds
    to $e^{-\lambda^{(2)} t}$ decay, with $\lambda^{(2)}$ given by
    Eq.~(\ref{2.33}).}
\label{fig-9}
\end{figure}

Figure~\ref{fig-9} shows peaks of the wave packet autocorrelation
function for the three-disk billiard with $\alpha=\sqrt{2}$,
$\beta=2$, $a=1$ and $R=10^4$. The peaks fall inside a narrow
cone. Magnitudes of the relatively strongest recurrence peaks decay
exponentially with time as $e^{-\lambda_\beta^{(2)} t}$, with the
decay rate $\lambda_\beta^{(2)}$ predicted by Eq.~(\ref{3.32}). The
trend of this exponential decay is shown by the solid line.  The
dashed line represents the exponential decay $e^{-\lambda^{(2)} t}$
due to the shortest two-disk periodic orbit in the system. The value
of $\lambda^{(2)}$ is calculated in accordance with Eq.~(\ref{2.33}),
and represents the fastest decay rate in the billiard.

As shown in Section II, the peaks of the autocorrelation function have
significant width which increases with time, see figs.~\ref{fig-2},
\ref{fig-4} and \ref{fig-6}. Therefore, if the autocorrelation
function peaks are closely spaced, as in fig.~\ref{fig-9}, the peak
broadening ultimately results in overlapping of neighboring peaks, so
that only the overall envelope decay $C(t)\sim e^{\lambda_\beta^{(2)} t}$ can be
resolved.  Thus, the simple semiclassical approach of the section
allows one to predict main features of the time-dependent
autocorrelation function for wave packets in arbitrary shaped
three-disk billiards.

\bigskip

Finally, let us compare the decays of the wave packet autocorrelation
function in three-disk scattering systems of three possible types: (i)
equilateral ($\alpha=\beta=1$), (ii) isosceles ($\alpha=\beta\neq1$)
and (iii) generic ($\alpha\neq\beta\neq1$) three-disk billiards.  Let
also the triangles, constituted by the disks centers, have
approximately equal side lengths, and differ only by the number of
symmetries. Thus, in the cases (ii) and (iii) both $\alpha$ and
$\beta$ are close (but not identical) to unity.
\begin{table}[h]
\begin{center}
\begin{tabular}{|c|c|}
  \hline
  \: Disk billiard \: & \: Decay exponent \: \\
  \hline
  $\begin{array}{c}
    \mathrm{Equilateral} \\
    \alpha = \beta = 1
  \end{array}$ & $\dfrac{v}{R} \ln \dfrac{0.54 \, R}{a}$\\
  \hline
  $\begin{array}{c} \mathrm{Isosceles} \\ \alpha = \beta \neq 1
  \end{array}$ & $\dfrac{v}{R} \ln \dfrac{1.04 \, R}{a}$\\
  \hline
  $\begin{array}{c}
    \mathrm{Generic} \\
    \alpha \neq \beta \neq 1
  \end{array}$ & $\dfrac{v}{R} \ln \dfrac{2R}{a}$\\
  \hline
\end{tabular}
\end{center}
\caption{Autocorrelation function decay exponents for three-disk
  billiards of different symmetries. All billiard are based on almost
  equilateral triangles, i.e. $\alpha$ and $\beta$ in the second and
  third row of the table are close to unity.}
\label{tab-1}
\end{table}
Table~\ref{tab-1} represents the autocorrelation function decay
exponents for the above-mentioned three-disk billiards calculated in
accordance with Eqs.~(\ref{2.43}), (\ref{2.51}) and (\ref{3.32}). One
can see that the decay rate increases by approximately $(v/R)\ln \! 2$
as the number of equal sides in the three-disk billiard decreases by
one. This difference in the decay rates is approximately equal to
twice the difference in KS-entropies per unit time of corresponding
billiards, and will be discussed in more details in the sequel.

Presence of symmetries in a scattering system increases the number of
periodic trajectories of a given length, and thus enhances
interference effects. It is due to the interference that strong wave
packet reconstruction peaks occur resulting in slower envelope decay
of the wave packet autocorrelation function.

\subsubsection{Hard-sphere billiards in three dimensions}

The above semiclassical approach can also be applied to calculate the
autocorrelation function decay for wave packets in three-dimensional
hard-sphere billiards. Here we briefly derive the autocorrelation
function decay rates for the following three systems: (i) the
two-sphere, (ii) the three-sphere equilateral, and (iii) the
four-sphere tetrahedral
\footnote{The tetrahedron is a triangular pyramid having congruent
equilateral triangles for each of its faces.} billiards.

The wave packet (phase space distribution) partial reconstruction
peaks of the semiclassical (classical) autocorrelation function are
given by Eq.~(\ref{3.8}) (Eq.~(\ref{3.9})), with the collision
sequences $\eta_{\gamma'}$ satisfying Eq.~(\ref{3.11}). The
probability weight $D_{\gamma'}$ of the periodic orbits $\gamma'$ are
now determined from Eq.~(\ref{3.10}) with $d=3$, and with the
differential cross section
\begin{equation}
\sdif(\theta) = \frac{a^2}{4},
\label{3.33}
\end{equation}
where $a$ stands for the radius of the hard-sphere scatterer. It will
become clear below that the independence of the differential cross
section $\sdif$ on the scattering angle $\theta$ significantly
simplifies the calculation of the decay rates of $C(t)$ and
$C_\mathrm{cl}(t)$.

\bigskip

Let us start with analyzing the autocorrelation function peaks for a
wave packet moving in a two-sphere billiard consisting of two hard
spheres, ``1'' and ``2'', of radius $a$, separated by a distance $R\gg
a$, e.g. see fig.~\ref{fig-1}. A wave packet is initially located on
the line connecting the sphere centers and moves toward one of the
spheres. As in the two-disk billiard case, there exists only one
periodic collision sequence, ``1212\ldots'', contributing to the
partial wave packet (phase space distribution) reconstruction at times
$t_n=2n R/v$, with $n=1,2,\ldots$. Hence,
\begin{equation}
C(t_n) \sim C_\mathrm{cl}(t_n) \sim \left( \frac{a^2}{4 R^2}
\right)^{2n} = \exp \left( -\gamma_\mathrm{3D}^{(2)} t_n \right),
\label{3.34}
\end{equation}
where
\begin{equation}
\gamma_\mathrm{3D}^{(2)} = \gamma_\mathrm{cl,3D}^{(2)} =
2\lambda^{(2)} = \frac{2v}{R} \ln \frac{2R}{a}
\label{3.35}
\end{equation}
is the sum of the two positive Lyapunov exponents,
$\lambda^{(2)}=(v/R)\ln(2R/a)$, of the two-sphere periodic orbit. As
in the two-disk billiard case, the absence of the interference of
different trajectories results in equality of the semiclassical and
classical decay rates.

\bigskip

The equilateral three-sphere billiard in three dimensions consists of
three spheres of radius $a$ placed in the vertices of an equilateral
triangle with side $R\gg a$, e.g. see fig.~\ref{fig-3}. In this
system, the peaks of the autocorrelation function occur at times
$t_n=n R/v$, with $n=2,3,4,\ldots$ counting the number of
collisions. The probability amplitude of a closed paths $\gamma'$,
given by $\sqrt{D_{\gamma'}(t_n)}=(a/2R)^n$, is only a function of
time $t_n$ and does not depend on the details of the particular
collision sequence $\eta_{\gamma'}=\{i,j,\ldots,q,r,s\}$ constituting
the orbit $\gamma'$. The number $M(t_n)$ of all collision sequences
$\eta_{\gamma'}$ satisfying Eq.~(\ref{3.11}) for $t=t_n$, and
therefore the number of interfering trajectories, can be calculated
with the help of the three-dimensional transition matrix
\begin{equation}
\vq_\mathrm{3D} = 
\begin{array}{cc}
  \!\! ^{1\cdot 2} \; ^{1\cdot 3} \; ^{2\cdot 1} \; ^{2\cdot 3}
  \; ^{3\cdot 1} \; ^{3\cdot 2} & \\ \left(
\begin{array}{cccccc}
0\; & 0\; & 1\; & 1\; & 0\; & 0\;\\ 0\; & 0\; & 0\; & 0\; & 1\; &
1\;\\ 1\; & 1\; & 0\; & 0\; & 0\; & 0\;\\ 0\; & 0\; & 0\; & 0\; & 1\;
& 1\;\\ 1\; & 1\; & 0\; & 0\; & 0\; & 0\;\\ 0\; & 0\; & 1\; & 1\; &
0\; & 0\; \end{array} \right) &
\begin{array}{c}
\!\! ^{1\cdot 2}\\
\!\! ^{1\cdot 3}\\
\!\! ^{2\cdot 1}\\
\!\! ^{2\cdot 3}\\
\!\! ^{3\cdot 1}\\
\!\! ^{3\cdot 2}
\end{array}
\end{array}
\label{3.36}
\end{equation}
in accordance with
\begin{equation}
M(t_n) = \left( \vq_\mathrm{3D}^n \right)_{1,1}.
\label{3.37}
\end{equation}
Here, as above, the subscript ``1,1'' denotes that the one-one element
of the matrix is taken. Then, the magnitude of the autocorrelation
function peak at time $t_n$ is given by $C(t_n) \sim [M(t_n)
(a/2R)^n]^2$. For long times, $n\gg 1$, the decrease of the peak
strength due to one collision can be written as
\begin{equation}
\frac{C(t_n+R/v)}{C(t_n)} = \left( \frac{\left( \vq_\mathrm{3D}^{n+1}
\right)_{1,1}}{\left( \vq_\mathrm{3D}^n \right)_{1,1}} \right)^2
\left( \frac{a}{2R} \right)^2 \stackrel{n\to \infty}{\longrightarrow}
\: \left( \frac{a}{R} \right)^2.
\label{3.38}
\end{equation}
Here we used that for $n\gg 1$ the ratio of the matrix elements in the
last equation is given by the largest eigenvalue of the matrix
$\vq_\mathrm{3D}$, which is equal to $2$. Therefore, we have
\begin{equation}
C(t_n) \sim \exp \left( -\gamma_\mathrm{3D}^{(3)} \: t_n \right),
\label{3.39}
\end{equation}
where
\begin{equation}
\gamma_\mathrm{3D}^{(3)} = \frac{2v}{R} \ln \frac{R}{a}
\label{3.40}
\end{equation}
is the autocorrelation function decay rate for the three-sphere
equilateral billiard in three dimensions.

The classical decay rate is calculated in the analogous way. According
to Eq.~(\ref{3.9}) we write $C_\mathrm{cl}(t_n) \sim M(t_n)
(a/2R)^{2n}$. Then,
\begin{equation}
\frac{C_\mathrm{cl}(t_n+R/v)}{C_\mathrm{cl}(t_n)} = \frac{\left(
\vq_\mathrm{3D}^{n+1} \right)_{1,1}}{\left( \vq_\mathrm{3D}^n
\right)_{1,1}} \left( \frac{a}{2R} \right)^2 \stackrel{n\to
\infty}{\longrightarrow} \: \left( \frac{a}{\sqrt{2} R} \right)^2,
\label{3.41}
\end{equation}
and consequently
\begin{equation}
C_\mathrm{cl}(t_n) \sim \exp \left( -\gamma_\mathrm{cl,3D}^{(3)} \:
t_n \right),
\label{3.42}
\end{equation}
where
\begin{equation}
\gamma_\mathrm{cl,3D}^{(3)} = \frac{2v}{R} \ln \frac{\sqrt{2} R}{a}
\label{3.43}
\end{equation}
is the classical autocorrelation function decay rate of the
three-sphere equilateral billiard. Below we will see that the
difference between $\gamma_\mathrm{cl,3D}^{(3)}$ and $\gamma_\mathrm{3D}^{(3)}$
is equal to the topological entropy per unit time (which in this
particular system coincides with the KS-entropy per unit time) of the
chaotic repeller of the classical system.

\bigskip

\begin{figure}[h]
  \centerline{\epsfig{figure=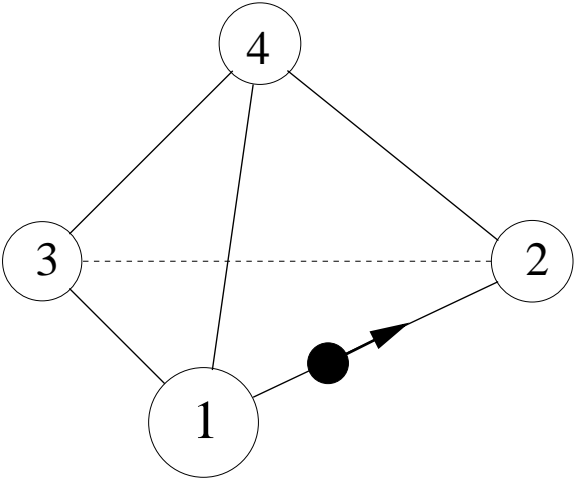,width=3.0in}}
  \caption{The tetrahedral four-sphere billiard. The wave packet is
    initially placed between spheres ``1'' and ``2'', with its average
    momentum directed toward sphere ``2''.}
\label{fig-10}
\end{figure}

Finally, we consider a substantially three-dimensional scattering
system -- a tetrahedral billiard -- where four spheres of radius $a$
are placed in the vertices of a pyramid build of four equilateral
triangles of sides $R\gg a$. The wave packet starts on a line
connecting two disks labeled by ``1'' and ``2'', see
fig.~\ref{fig-10}.

The calculation of the autocorrelation function decay exponent for
this billiard proceeds in close analogy with the three-sphere case
considered above. The transition matrix $\tilde{\vq}_\mathrm{3D}$ is
now a $12\times12$ matrix connecting directions ($1\to 2$), ($1\to
3$), ($1\to 4$), ($2\to 1$), ($2\to 3$), ($2\to 4$), ($3\to 1$),
($3\to 2$), ($3\to 4$), ($4\to 1$), ($4\to 2$) and ($4\to 3$). The
structure of the matrix is similar to the one given by
Eq.~(\ref{3.36}) for the case of the three-sphere billiard; the largest
eigenvalue is now equal to $3$. As before the number of periodic
trajectories contributing to the autocorrelation peak at time $t_n = n
R/v$, $n=2,3,\ldots$, is given by Eq.~(\ref{3.37}) with
$\tilde{\vq}_\mathrm{3D}$ being the $12\times12$ transition matrix of
the four-sphere tetrahedral billiard. Then, the decrease of the
autocorrelation function peak strength due to one collision is given
by
\begin{equation}
\frac{C(t_n+R/v)}{C(t_n)} = \left( \frac{\left( \vq_\mathrm{3D}^{n+1}
\right)_{1,1}}{\left( \vq_\mathrm{3D}^n \right)_{1,1}} \right)^2
\left( \frac{a}{2R} \right)^2 \stackrel{n\to \infty}{\longrightarrow}
\: \left( \frac{3 a}{2 R} \right)^2,
\label{3.44}
\end{equation}
leading to
\begin{equation}
C(t_n) \sim \exp \left( -\gamma_\mathrm{3D}^{(4)} \: t_n \right),
\label{3.45}
\end{equation}
with the four-sphere escape decay rate given by
\begin{equation}
\gamma_\mathrm{3D}^{(4)} = \frac{2v}{R} \ln \frac{2 R}{3 a}.
\label{3.46}
\end{equation}

The classical autocorrelation function decay rate for the four-sphere
tetrahedral billiard is obtained straightforwardly. According to
Eq.~(\ref{3.9}) we write
\begin{equation}
\frac{C_\mathrm{cl}(t_n+R/v)}{C_\mathrm{cl}(t_n)} = \frac{\left(
\vq_\mathrm{3D}^{n+1} \right)_{1,1}}{\left( \vq_\mathrm{3D}^n
\right)_{1,1}} \left( \frac{a}{2R} \right)^2 \stackrel{n\to
\infty}{\longrightarrow} \: \left( \frac{\sqrt{3} a}{2 R} \right)^2.
\label{3.47}
\end{equation}
This results to
\begin{equation}
C_\mathrm{cl}(t_n) \sim \exp \left( -\gamma_\mathrm{cl,3D}^{(4)} \:
t_n \right),
\label{3.48}
\end{equation}
with
\begin{equation}
\gamma_\mathrm{cl,3D}^{(4)} = \frac{2v}{R} \ln \frac{2 R}{\sqrt{3} a}.
\label{3.49}
\end{equation}
The semiclassical decay rate, $\gamma_\mathrm{3D}^{(4)}$, is again
found to be slower than the classical one,
$\gamma_\mathrm{cl,3D}^{(4)}$. As we will show below the difference
between the two decay rates is given by the topological entropy per
unit time of the four-sphere tetrahedral billiard repeller.

\bigskip

\begin{figure}[h]
  \centerline{\epsfig{figure=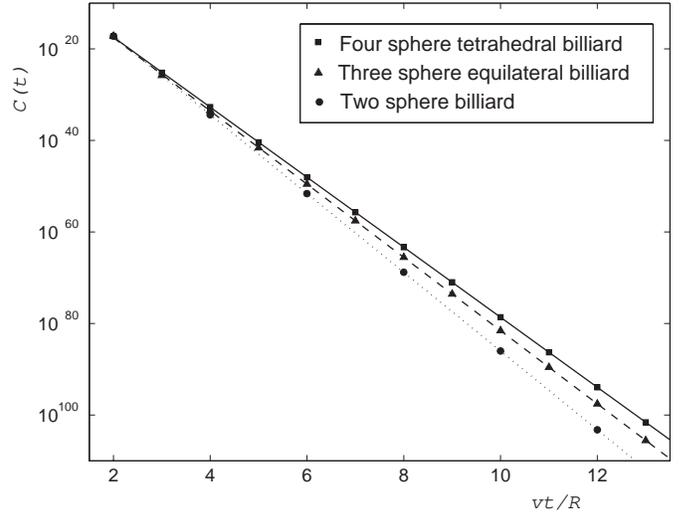,width=3.5in}}
  \caption{Peaks of the wave packet autocorrelation function for the
    three scattering systems: two-sphere (circles), three-sphere
    equilateral (triangles) and four-sphere tetrahedral (diamonds)
    billiards. In all cases the radius of the sphere scatterers is
    $a=1$, and the sphere center-to-center separation is $R=10^4$. The
    semiclassical decay rates for these billiards are calculated
    according to Eqs.~(\ref{3.35}), (\ref{3.40}) and (\ref{3.46}), and
    presented in the figure by the dotted, dashed and solid lines
    respectively.}
\label{fig-11}
\end{figure}

Figure \ref{fig-11} shows relative magnitude of peaks of the wave
packet autocorrelation function for the three scattering systems
considered above: two-sphere, three-sphere equilateral and four-sphere
tetrahedral billiards. The radius of the sphere scatterers is $a=1$,
and the sphere center-to-center separation is $R=10^4$. The
semiclassical autocorrelation function decay rates for these billiards
are calculated according to Eqs.~(\ref{3.35}), (\ref{3.40}) and
(\ref{3.46}), and presented in the figure by the dotted, dashed and
solid lines respectively. The figure illustrates the reduction of
escape of a semiclassical particle from a billiard as additional
scatterers are added to the latter. Below we clarify the connection of
the rate of this escape to classical properties of the system's
chaotic repeller.

\subsection{Classical and semiclassical escape rates}

We would now like to comment on the connection of the escapes rates in
classical and semiclassical open billiards to such properties of
classical chaotic systems as the mean Lyapunov exponents, the
Kolmogorov-Sinai (KS) and topological entropy.

The classical escape rate $\gamma_\mathrm{cl}$, which is equivalent to the
decay rate of the classical autocorrelation function, is known to
equal a difference of the stretching and randomization rates in a
chaotic system \cite{gasp-book,kantz,dorf},
\begin{equation}
\gamma_\mathrm{cl} = \sum_{\lambda_i>0} \lambda_i - h_\mathrm{KS}.
\label{3.50}
\end{equation}
Here, the sum, going over all mean positive Lyapunov exponents $\lambda_i$,
represents the rate of the local exponential stretching of an initial
particle phase space distribution, and $h_\mathrm{KS}$ is the
KS-entropy per unit time of the system characterizing the rate at
which the phase space distribution gets randomized over the chaotic
repeller of the scattering system. It was first pointed out by Gaspard
and Rice \cite{gasp-II} that the semiclassical escape rate $\gamma$ for the
three hard-disk scattering system in two dimensions is given by
\begin{equation}
\gamma \approx \lambda - 2h_\mathrm{KS},
\label{3.51}
\end{equation}
where $\lambda$ is the mean positive Lyapunov exponent of the corresponding
chaotic repeller. Comparison of Eqs.~(\ref{3.50}) and (\ref{3.51})
shows that the semiclassical escape rate is never greater than the
escape rate in the counterpart classical system, which is consistent
with our observations earlier in this section.

It is interesting to note that the equality in Eq.~(\ref{3.51}) can be
extended (and made exact!) for the three-dimensional hard-sphere
billiards considered in this section, i.e. for the two-sphere,
three-sphere equilateral and four-sphere tetrahedral scattering
systems. Moreover, Eq.~(\ref{3.51}) can be easily derived for these
billiards, since the KS-entropy of the repellers in these systems has
an especially simple form. Indeed, the independence of the hard-sphere
differential cross section on the scattering angle, evident from
Eq.~(\ref{3.33}), and the equality of lengths of all free flight
segments composing the system's periodic orbit, result in equivalence
of probability weights for all trajectories of given length
constituting the chaotic repeller, see Eq.~(\ref{3.10}). Since the
probability measure is the same for all periodic trajectories
involving a given number of collisions, the construction of the
KS-entropy is identical to the construction of the topological
entropy, e.g. see \cite{dorf}.  Therefore, for the particular
three-dimensional billiards of this section, the KS-entropy per unit
time, $h_\mathrm{KS}$, of the repeller is simply equal to the
topological entropy per unit time, $h_\mathrm{top}$. The later is the
rate of the exponential growth with time $t$ of the number of
different collision sequences, $M(t)$, a classical particle moving on
the repeller can possibly undergo:
\begin{equation}
M(t) \sim \exp(h_\mathrm{top} \, t).
\label{3.52}
\end{equation}

For the two-sphere billiard there is only one periodic collision
sequence a particle staying on the repeller can follow. Thus, $M(t)=
1$ and
\begin{equation}
h_\mathrm{top}^{(2)} = 0.
\label{3.53}
\end{equation}
In the three-sphere equilateral billiard the number of possible
trajectories multiplies by two at every collision, so that $M(t) =
2^{vt/R}$, leading to
\begin{equation}
h_\mathrm{top}^{(3)} = \frac{v}{R} \ln 2.
\label{3.54}
\end{equation}
In the same way we conclude that in the four-sphere tetrahedral
billiard the number of possible collision sequences is $M(t) =
3^{vt/R}$, so that the topological entropy per unit time reads
\begin{equation}
h_\mathrm{top}^{(4)} = \frac{v}{R} \ln 3.
\label{3.55}
\end{equation}

Every trajectory $\gamma'$ of the three-dimensional hard-sphere
billiard repeller has two positive Lyapunov exponents,
$\lambda_1(\gamma')$ and $\lambda_2(\gamma')$, corresponding to
perturbations of the particle's initial conditions in two direction
perpendicular to the trajectory. It is straightforward to show that
the sum of the two Lyapunov exponent, for a dilute hard-sphere
scattering system, $R\gg a$, does not depend on the details of the
trajectory $\gamma'$, and is given by \cite{disser}
\begin{equation}
\lambda_1 + \lambda_2 = \frac{2v}{R} \ln\frac{2R}{a}.
\label{3.56}
\end{equation}
The last equality is obvious for the two-sphere periodic orbit, since
in that case $\lambda_1=\lambda_2=\lambda^{(2)}=(v/R)\ln(2R/a)$. In
the case of a general hard-sphere dilute scattering system the method
of curvature radii \cite{us,gasp-book} can be used to prove
Eq.~(\ref{3.56}).

Looking at the expressions for the classical autocorrelation function
decay rates in the three-dimensional billiards studied above,
Eqs.~(\ref{3.35}), (\ref{3.43}) and (\ref{3.49}), we notice that the
following relation holds
\begin{equation}
\gamma_\mathrm{cl,3D}^{(j)} = \lambda_1 + \lambda_2 -
h_\mathrm{top}^{(j)},
\label{3.57}
\end{equation}
with $j=2,3,4$. One can also verify that the semiclassical decay
rates, given by Eqs.~(\ref{3.35}), (\ref{3.40}) and (\ref{3.46}),
satisfy
\begin{equation}
\gamma_\mathrm{3D}^{(j)} = \lambda_1 + \lambda_2 - 2
h_\mathrm{top}^{(j)}.
\label{3.58}
\end{equation}
Equations~(\ref{3.57}) and (\ref{3.58}) are the three-dimensional
version of Eqs.~(\ref{3.50}) and (\ref{3.51}) respectively.

The appearance of the factor of $2$ in the expression for the
semiclassical decay rate is now apparent: the strength of the wave
packet partial reconstruction peaks, $C(t)$, is proportional to
$M^2(t)$ due to interference of different periodic orbits, while in
the classical case $C_\mathrm{cl}(t)$ is given merely by the sum of
orbit probabilities, and is proportional to $M(t)$. One can also see
now that the difference of the two escape rates is given by the
topological entropy (which, for the three-dimensional billiards in
question, is equal to the KS-entropy) of the chaotic repeller of the
classical system, and is therefore determined by underlying classical
dynamics.

\bigskip

\begin{figure}[h]
  \centerline{\epsfig{figure=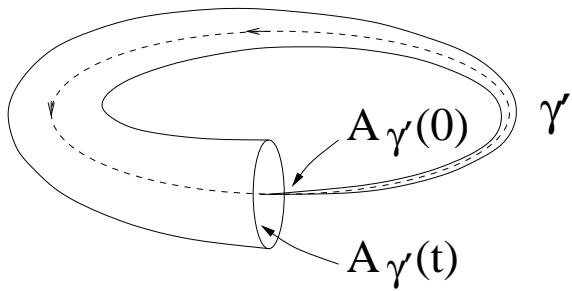,width=3.0in}}
  \caption{Schematic picture of a classical periodic orbit $\gamma'$
    together with an exponentially spreading tube of infinitesimally
    close trajectories. The initial cross section of the tube,
    $A_{\gamma'}(0)$, gets magnified to the value $A_{\gamma'}(t)$, given by
    Eq.~(\ref{3.59}), after time $t$ equal to the period of the orbit
    $\gamma'$.}
\label{fig-12}
\end{figure}

Finally, we would like to further clarify the appearance of the factor
of $2$ in the expression for the semiclassical escape rate,
Eq.~(\ref{3.58}), by means of the following simple argument. Consider
a classical periodic orbit $\gamma'$ of the period $t$, schematically
depicted in fig.~\ref{fig-12}, passing through the phase space point
around which the initial wave packet is localized. In a general
chaotic system the number of such periodic orbits $M(t)$ grows
exponentially with the period $t$, and the rate of this growth is
given by the topological entropy per unit time, $h_\mathrm{top}$, see
Eq.~(\ref{3.52}). In order to determine the probability weight
$D_{\gamma'}(t)$ of the orbit $\gamma'$, one needs to consider a
bundle of trajectories which stay infinitesimally close to
$\gamma'$. These trajectories form an exponentially thickening tube,
see fig.~\ref{fig-12}, with the initial and final cross sections,
$A_{\gamma'}(0)$ and $A_{\gamma'}(t)$ respectively, related by
\begin{equation}
A_{\gamma'}(t) = A_{\gamma'}(0) \exp\left( \sum_{\lambda_j(\gamma') > 0}
\lambda_j(\gamma') \: t \right).
\label{3.59}
\end{equation}
Here the sum goes over all positive Lyapunov exponents
$\lambda_j(\gamma')$ of the periodic orbit $\gamma'$. Then, the
probability for a classical particle taken at random from the
trajectory bundle to stay on the periodic orbit $\gamma'$ is
proportional to the ratio $A_{\gamma'}(0)/A_{\gamma'}(t)$, so that
\begin{equation}
D_{\gamma'}(t) \sim \exp\left( -\sum_{\lambda_j(\gamma') > 0}
\lambda_j(\gamma') \: t \right).
\label{3.60}
\end{equation}
Equation~(\ref{3.60}) together with (\ref{3.52}) leads to the
expression for the semiclassical escape rate in the case when the sum
over positive Lyapunov exponents, $\sum \lambda_j(\gamma')$, does not significantly
depend on the details of the orbit $\gamma'$, and can be replaced by its
average value $\sum \lambda_j$. Then, the sum over orbits $\{ \gamma'\}$ in
Eq.~(\ref{3.8}) is simply equal to the product of the number of these
orbits, $M(t)$, given by Eq.~(\ref{3.52}), and the probability
amplitude, $\sqrt{D(t)}$, identical for all the orbits and calculated
in accordance with Eq.~(\ref{3.60}). The substitution yields
\begin{equation}
C(t) \sim \left( M(t) \sqrt{D(t)} \right)^2 \sim \exp\left[ -\left(
\sum_{\lambda_j > 0} \lambda_j - 2 h_\mathrm{top} \right) \right].
\label{3.61}
\end{equation}
Thus, we recover the expression for the semiclassical autocorrelation
function decay rate, Eq.~(\ref{3.58}), obtained for a number of
hard-sphere billiards in three spatial dimensions.

The purpose of the above oversimplified derivation is only to
illustrate the origin of the intimate relation between the quantum
escape rate and such properties of the underlying classical chaotic
system as the Lyapunov exponents and topological (and/or KS) entropy.
The details of this relation as well as the limits of its
applicability are yet to be investigated.

\section{Summary and conclusions}

In the first part of this paper we used the technique of multiple
collision expansions to construct the quantum propagator for a
particle with the de Broglie wavelength, $\la$, traveling in an array
of hard-disk scatterers of radius $a\gg \la$. The scattering system was
assumed to be so dilute that the typical scatterer separation $R$
satisfied the condition $R\gg a^2/\la$. The quantum propagator was used
to analytically calculate the time-dependent autocorrelation function
for a wave packet, initially localized in both position and momentum
spaces, evolving in open two- and three-disk billiard systems. It was
found that the autocorrelation function exhibits a sequence of sharp
peaks at times multiple to periods of classical phase space periodic
orbits of the billiards. These peaks correspond to partial
reconstructions of the initial wave packet in the course of its time
evolution. The envelope of the correlation function decays
exponentially with time after one or two particle-disk collisions; the
exponential decay lasts for some $a/ \la\gg 1$ scattering events. Our
calculations recovered the autocorrelation function decay rate, first
obtained in reference \cite{gasp-II}, for a particle moving in the
three-disk equilateral billiard, and predicted the detailed structure
and the decay rate of the autocorrelation function for the more
complicated three-disk isosceles billiard.

In the second part of this paper the method of the semiclassical Van
Vleck propagator was utilized to derive a simple expression for the
relative magnitude of the wave packet partial reconstruction peaks.
Although fails to describe the full time dependence of the
autocorrelation function, this method allows one to calculate,
analytically or numerically, the decay rates of the autocorrelation
function envelope (also known as particle escape rates) for much more
complicated hard-disk and hard-sphere scattering systems. Here we used
it to analyze the three-disk generic billiard, the two-sphere,
three-sphere equilateral and three-sphere tetrahedral scattering
systems.

A straightforward modification of the semiclassical method allowed us
to construct the expression for the peaks of the classical
autocorrelation function, and to compare classical and semiclassical
particle escape rates in various open hard-disk and hard-sphere
billiards. The semiclassical escape rate in three-dimensional
scattering systems was shown to be given by the difference of the sum
of the two positive Lyapunov exponents and twice the topological
entropy (in this case equal to the KS-entropy) per unit time of the
underlying classical repeller. Thus, the semiclassical escape rate is
never grater then the classical one, and the difference between the
two equals to the topological (or KS) entropy per unit time of the
classical system. This result is consistent with the earlier findings
of reference \cite{gasp-II} for the case of the two-dimensional
hard-disk billiards, and therefore strengthens the connection between
classical and quantum chaos.

An interesting seeming ``paradox'' arises if one compares the
expressions for the classical and semiclassical escape rates, e.g. see
Eqs.~(\ref{3.57}) and (\ref{3.58}). Indeed, according to the
Correspondence Principle it is expected that in the limit of the de
Broglie wavelength going to zero the semiclassical escape rate should
match its classical counterpart. This seems to be incompatible with
the results of this paper: the semiclassical escape rate is expressed
in terms of classical quantities only (and does not depend on the de
Broglie wavelength), and (in the case of hard-sphere billiards) is
smaller than the classical one by the value of the topological entropy
of the system. In fact, the difference of the two escape rates
originates from noncommutivity of the order in which the infinite time
limit and the classical (de Broglie wavelength going to zero) limit
are taken. This question has been carefully studied by Barra and
Gaspard for the case of quantum graphs \cite{barra}, as well as
discussed in Ref.  \cite{gasp-IV}.

\section*{Acknowledgments}

The authors would like to thank Pierre Gaspard, Shmuel Fishman,
Christopher Jarzynski, Fernando Cucchietti, Sudhir Jain and Arnd
B\"acker for helpful conversations.  J.R.D wishes to thank the National
Science Foundation for support under Grant No. PHY-01-38697. A.G.
would like to acknowledge the Alexander von Humboldt Foundation
(Germany) for support at the final stage of the research.

\appendix
\section{Circular wave packet expansion}

Here we derive the expansion of the circular wave packet given by
Eqs.~(\ref{2.14}) and (\ref{2.15}). The function $\chi_l(k,k_0)$, defined
in Eq.~(\ref{2.15}), can be put in the integral form \cite{gradst}
\begin{equation}
  \chi_l(k,k_0) = \frac{2\sqrt{\pi}}{\sigma} \int_0^\sigma dr r J_l(kr) J_l(k_0 r).
\label{A1}
\end{equation}
Then,
\begin{equation}
\begin{split}
  \sum_{l=-\infty}^{+\infty} &\chi_l(k,k_0) e^{i l (\theta_k - \theta_{k_0})}\\ &=
  \frac{2\sqrt{\pi}}{\sigma}
  \int_0^\sigma dr r \sum_{l=-\infty}^{+\infty} J_l(kr) J_l(k_0 r) e^{i l (\theta_k - \theta_{k_0})}\\
  &= \frac{2\sqrt{\pi}}{\sigma} \int_0^\sigma dr r J_0(|\vk-\vk_0|r),
\end{split}
\label{A2}
\end{equation}
where the ``summation theorem'' for zeroth order Bessel function
\cite{gradst} was used. Doing the simple integral we end up with the
equality
\begin{equation}
  \sum_{l=-\infty}^{+\infty} \chi_l(k,k_0) e^{i l (\theta_k - \theta_{k_0})} 
  = 2\sqrt{\pi} \frac{J_0(|\vk-\vk_0|\sigma)}{|\vk-\vk_0|}.
\label{A3}
\end{equation}

\end{document}